\documentclass[review,english]{elsarticle}

\usepackage{lineno,hyperref}
\modulolinenumbers[5]

\journal{Medical Image Analysis}
\biboptions{authoryear}
\usepackage[T1]{fontenc}
\usepackage[latin1]{inputenc}
\usepackage[percent]{overpic}
\usepackage{pict2e}
\usepackage{graphicx}
\usepackage{booktabs}
\usepackage{threeparttable}
\usepackage{float}
\usepackage{csquotes}

\usepackage[usenames,dvipsnames,table]{xcolor}

\usepackage{soul}

\usepackage{graphicx}
\usepackage{psfrag}
\usepackage{epsfig,subfigure}
\usepackage{epstopdf}
\expandafter\let\csname equation*\endcsname\relax
\expandafter\let\csname endequation*\endcsname\relax
\usepackage{algorithm}
\usepackage{algcompatible}

\usepackage{amsmath}
\mathchardef\mhyphen="2D

\usepackage{babel}
\makeatother
\usepackage{url}
\begin{document}

\begin{frontmatter}

\title{Directional-TV Algorithm for Image Reconstruction from Limited-Angular-Range Data}

\author[a]{Zheng Zhang}
\author[a]{Buxin Chen}
\author[a]{Dan Xia}
\author[a]{Emil Y. Sidky}
\author[a,b]{Xiaochuan Pan\fnref{myfootnote}}
\address[a]{Department of Radiology, The University of Chicago, Chicago, IL 60637, USA}
\address[b]{Department of Radiation and Cellular Oncology, The University of Chicago, Chicago, IL 60637, USA}
\fntext[myfootnote]{Corresponding author. Email: xpan@uchicago.edu.}

\begin{abstract}

Investigation of image reconstruction from data collected over a limited-angular range in X-ray CT remains a topic of active research because it may yield insight into the development of imaging workflow of practical significance. This reconstruction problem is well-known to be challenging, however, because it is highly ill-conditioned. In the work, we investigate optimization-based image reconstruction from data acquired over a limited-angular range that is considerably smaller than the angular range in short-scan CT. We first formulate the reconstruction problem as a convex optimization program with directional total-variation (TV) constraints applied to the image, and then develop an iterative algorithm, referred to as the {\it directional-TV} (DTV) algorithm for image reconstruction through solving the optimization program. 
We use the DTV algorithm to reconstruct images from data collected over a variety of limited-angular ranges for 
breast and bar phantoms of clinical- and industrial-application relevance. 
The study demonstrates that the DTV algorithm accurately recovers the phantoms from data generated over a significantly reduced angular range, and that it considerably diminishes artifacts observed otherwise in reconstructions of existing algorithms. We have also obtained empirical conditions on minimal-angular ranges sufficient for numerically accurate image reconstruction with the DTV algorithm. 
\end{abstract}

\begin{keyword}
limited-angle reconstruction \sep directional total variation \sep optimization-based reconstruction \sep primal-dual algorithm \sep computed tomography
\end{keyword}

\end{frontmatter}


\section{Introduction}
\label{sec:Intro}




 In X-ray CT imaging applications that arise in medicine and other fields, it is desirable often to collect data only over a limited-angular range  due to practical constraints on application workflows. Therefore, there remains a high level of interest in the field in investigating and developing appropriate image reconstruction from limited-angular-range data in CT imaging. The problem of image reconstruction from data collected over a limited-angular range (e.g., significantly smaller than $\pi$) is well-known to be challenging because it is highly ill-conditioned \citep{louis1986incomplete,frikel2013characterization, quinto2017artifacts}.

Iterative algorithms have been investigated with image constraints for potentially alleviating artifacts in reconstructions from limited-angular-range data, including constraints on image total variation (TV) \citep{Delaney1998, Sidky:06,jin2010anisotropic,wang2017reweighted}, on $\ell_0$-norm of image gradient \citep{yu2015}, with object's shape information \citep{schorr2013exploitation,liu2016cooperative}, and with discrete grey scales \citep{batenburg2011dart,zhuge2015tvr}.  In a recent work \citep{xu2018ct,xu2019image}, constraints were proposed on image's partial derivatives along two orthogonal directions for reducing limited-angular-range artifacts that are observed otherwise in reconstructions of existing analytic and iterative algorithms.

Accurate image reconstruction from data over limited-angular ranges remains  of theoretical and practical interest. 
In this work, we perform an investigation of image reconstruction from data acquired over a limited-angular range in CT by developing a new reconstruction algorithm, referred to as the directional TV (DTV) algorithm, and applying it to phantoms of distinct characteristics. Specifically, we first formulate the reconstruction problem as a convex optimization program in which a data divergence is minimized under two DTV constraints, and then develop an instance, i.e., the DTV algorithm, of the general primal-dual (PD) algorithm \citep{c-p:2011,sidky2012convex} to reconstruct an image through solving the optimization program.  We design explicit convergence conditions for monitoring and assessing the convergence characteristics of the DTV algorithm and for ensuring its numerically accurately solving the optimization program.  

We carry out numerical studies by using digital phantoms eyeing on two aims:  (1) to verify numerically the DTV algorithm and its computer implementation in image reconstruction from limited-angular-range data, and (2) to investigate  empirically {\it minimal-angular ranges} sufficient for numerically accurate image reconstruction permitted by the DTV algorithm.  In an attempt to investigate the impact of subject structure on reconstruction accuracy, we consider in our studies numerical breast \citep{jorgensen2012quantifying} and bar phantoms, as they are relevant to digital breast tomosynthesis (DBT)  \citep{niklason1997digital,dobbins2003digital,teuwen2020deep} and to industrial CT-imaging  \citep{de2014industrial,carmignato2018industrial} applications. We evaluate image reconstructions by using quantitative metrics for measurement of numerical and visual accuracy of reconstructions. The numerical studies with noiseless data can provide insights into the upper bound of reconstruction performance of the DTV algorithm. Additionally, we conduct a preliminary investigation of image reconstruction from data containing noise to demonstrate visually the noise impact on the DTV algorithm. 

Following the introduction in Sec. \ref{sec:Intro}, we develop the optimization program and DTV algorithm in Sec. \ref{sec:Alg}, and present numerical studies in Secs. \ref{sec:Breast-result} and \ref{sec:bar-result}. Discussion and conclusion are given in Secs. \ref{sec:Discussion} and \ref{sec:Conclusions}. We include the derivation and verification details of the DTV algorithm in Appendices A and B to avoid distracting the presentation flow in the main text.

\section{Methods}\label{sec:Alg}
We present the development of the DTV algorithm in the context of two-dimensional (2D) image reconstruction from fan-beam projection data over a limited-angular range. However, the work can readily be extended to a 3D cone-beam scanning configuration.

\subsection{Discrete X-ray transform (DXT)-data model}
\begin{figure}
\centering
\includegraphics[angle=0,trim=0 0 0 0, clip,origin=c,width=1\textwidth]{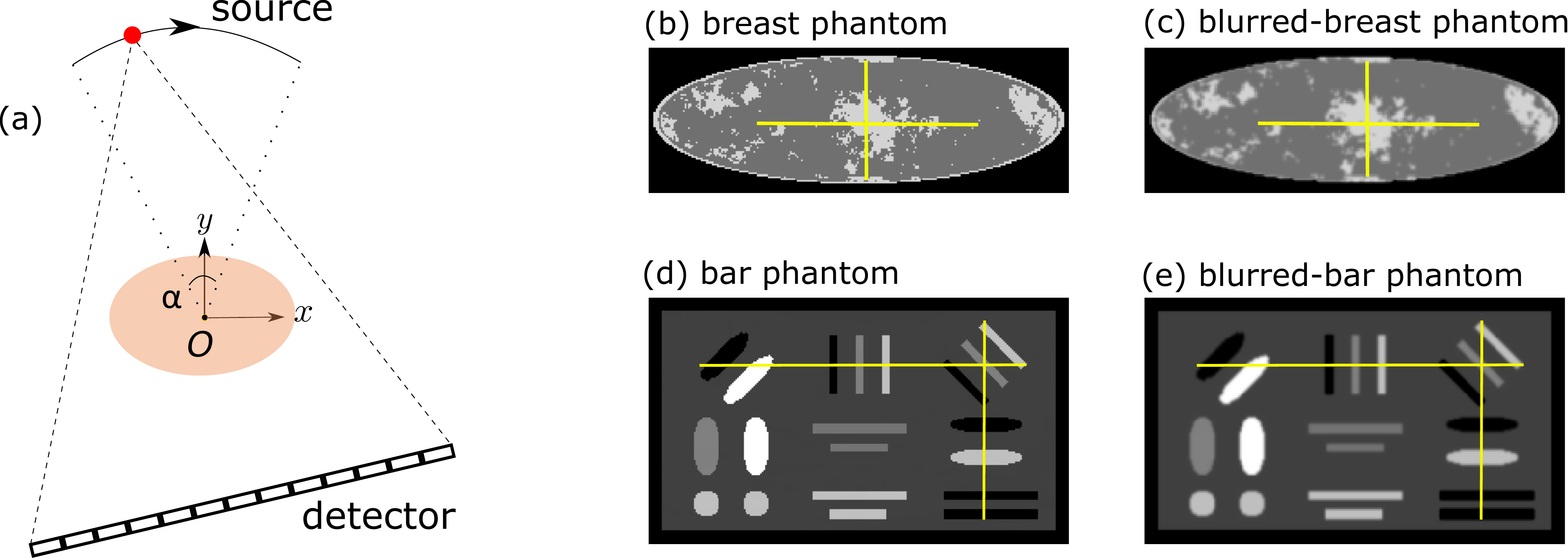}
\caption{(a) A scanning configuration collecting data over limited-angular range $\alpha$ with a pair of an X-ray source and a linear detector array; (b) and (c)  breast and blurred-breast phantoms; and (d) and (e) bar and blurred-bar phantoms. While the breast and bar phantoms are piece-wise constant, the blurred-breast and blurred-bar phantoms are not piece-wise constant. {  The horizontal and vertical yellow line segments over the phantoms indicate the loci over which image profiles are plotted in Secs. \ref{sec:Breast-result} and \ref{sec:bar-result} below}.}
\label{fig:config}
\end{figure}

In the scanning configuration considered, shown in Fig. \ref{fig:config}a, an X-ray source and a linear detector array simultaneously rotate around the center of rotation $O$. Data are collected over a circular arc, which forms limited-angular range $\alpha$, from an object that is within the field of view of the configuration. We assume that the circular arc is symmetric relative to $y$-axis, i.e., its starting and ending angles are $-\frac{\alpha}{2}$ and  $\frac{\alpha}{2}$.  We refer to a scan as a limited-angular-range scan if $\alpha$ is considerably less than the angular range of $\pi$ plus fan angle, i.e., a short-scan.  

Data collected in X-ray tomographic imaging with the scanning configuration in Fig. \ref{fig:config}a may be modeled by the discrete-to-discrete X-ray transform (DXT)-data model, which is given by 
\begin{equation}\label{eq:DXT}
\mathbf{g}(\mathbf{f})=\mathcal{H}\,\mathbf{f},
\end{equation}
where vector $\mathbf{g}$ of size $M=N_s \times N_d$ denotes model data, with entry $g_j$ the data element of ray $j$, where $j=0, 1, ..., M-1$; $N_s$ is the number of projection views, with angular interval between two adjacent views $\Delta \alpha = \frac{\alpha}{N_s - 1}$; $N_d$ is the number of detector bins of bin-size $\Delta_d$ of the linear detector array; 
vector $\mathbf{f}$ of size $N$ denotes a 2D discrete image, where $N=N_x \times N_y$, $N_x$ and $N_y$ the total number of pixels along $x$- and $y$-axis; $f_i$ depicts the $i$th entry of $\mathbf{f}$ on  the $i$th square-shaped pixel of size $\Delta_I$; and $\mathcal{H}$ depicts the system matrix of size $M\times N$ with element $h_{ji}$ representing the intersection length of ray $j$ with pixel $i$ \citep{siddon1985fast}.  While $\mathbf{g}$ is formed by concatenating data in the order of projection views, $\mathbf{f}$ is obtained by concatenating pixel values in the order of $x$- and $y$-axis. The breast and bar phantoms are discretized on image arrays of sizes $80\times 256$ and $150\times 256$, respectively. 

\subsection{Optimization program with DTV constraints}
Using  Eq. \eqref{eq:DXT}, we formulate the image-reconstruction problem as a convex optimization program:
\begin{equation}\label{eq:opt}
\mathbf{f}^{\star} = \underset{\mathbf{f}}{\mathsf{argmin}} D_{\mathbf{g}}(\mathbf{f}) \quad {\rm s.t.} \quad ||(|\mathcal{D}_x \mathbf{f}|)||_1 \le t_x, \,\, ||(|\mathcal{D}_y \mathbf{f}|)||_1 \le t_y, \,\, {\rm and} \,\, f_i \ge 0, 
\end{equation}
where $D_{\mathbf{g}}(\mathbf{f})$ is square of the $\ell_2$-norm of the difference between 
measured data  $\mathbf{g}^{[\mathcal{M}]}$ and model data $\mathbf{g}(\mathbf{f})$, i.e., 
\begin{equation}\label{eq:Dg-l2}
D_{\mathbf{g}}(\mathbf{f}) = \frac{1}{2} \parallel  \mathbf{g}^{[\mathcal{M}]} - \mathbf{g}(\mathbf{f}) \parallel_2^2;
\end{equation}
matrices $\mathcal{D}_x$ and $\mathcal{D}_y$ of size $N \times N$ denote two-point differences along $x$- and $y$-axis, respectively; vectors $\mathcal{D}_x \mathbf{f}$ and $\mathcal{D}_y \mathbf{f}$ are of size $N$ with elements given by
\begin{equation}
\begin{split}
(\mathcal{D}_x \mathbf{f})_k &= 
\begin{cases}
f_{k+1} - f_{k} \quad {\rm for} \,\,\,\, (k \bmod N_x) \ne N_x-1 \\
- f_{k} \hspace{1.2cm} {\rm for} \,\,\,\, (k \bmod N_x) = N_x-1
\end{cases} \\
(\mathcal{D}_y \mathbf{f})_l &= 
\begin{cases}
f_{l+N_x} - f_{l} \quad {\rm for} \,\,\,\, l < N-N_x  \\
- f_{l} \hspace{1.4cm} {\rm for} \,\,\,\, l \ge N-N_x;
\end{cases} \\
\end{split}
\end{equation}
scalars $||(|\mathcal{D}_x \mathbf{f}|)||_1$ and $||(|\mathcal{D}_y \mathbf{f}|)||_1$, referred to as the {\it directional TVs}, indicate the $\ell_1$-norms of image's partial derivatives along directions $x$ and $y$; and parameters $t_x$ and $t_y$ depict the upper bounds on the DTV constraints. The form of the optimization program, along with its parameters, including $t_x$ and $t_y$, specifies completely the solutions to the optimization program. The use of the DTV constraints is inspired by a recent work reported in Refs. \citep{xu2018ct,xu2019image}.

\subsection{Development and verification of the DTV algorithm}\label{sec:algorithm}


Our DTV algorithm is developed by basing upon the general primal-dual (PD) algorithm \citep{c-p:2011}, because the PD algorithm can solve accurately convex optimization programs such as the program in Eq. \eqref{eq:opt}.

\subsubsection{Development of the DTV algorithm}\label{sec:DTV}
The development of the DTV algorithm from the PD algorithm requires (a) to derive its two proximal-mapping problems tailored to the optimization program in Eq. \eqref{eq:opt} and (b) to solve the proximal-mapping problems derived \citep{sidky2012convex,sidky2014analysis,zhang2016artifact,xia2016optimization}. In \ref{sec:cpd}, we derive the DTV algorithm by tailoring the two proximal-mapping problems for Eq. \eqref{eq:opt}, and then by obtaining analytic solutions, which involve only a finite number of algebraic calculations, to the two proximal-mapping problems. Therefore, the DTV algorithm not only lends itself computational accuracy and efficiency, but also retains mathematical rigor of the PD algorithm.
For the sake of maintaining the presentation flow, we include the derivation details of the DTV algorithm in \ref{sec:cpd} and show  its pseudo-code in Algorithm \ref{alg:1}.
\begin{algorithm}[h]\leavevmode
\caption{Pseudo-code of the DTV algorithm for 
solving Eq. \eqref{eq:opt}}\label{alg:1}
\begin{algorithmic}[1]
\STATEx INPUT: $g^{[\mathcal{M}]}$, $t_x$, $t_y$, $\mathcal{H}$, $b$
\STATE $L \leftarrow ||\mathcal{K}||_2$, $\tau \leftarrow b/L$, $\sigma \leftarrow 1/(bL)$, $\nu_1 \leftarrow ||\mathcal{H}||_2/||\mathcal{D}_x||_2$, $\nu_2 \leftarrow ||\mathcal{H}||_2/||\mathcal{D}_y||_2$, $\mu \leftarrow ||\mathcal{H}||_2/||\mathcal{I}||_2$
\STATE $n \leftarrow 0$
\STATE INITIALIZE: $\mathbf{f}^{(0)}$, $\mathbf{w}^{(0)}$, $\mathbf{p}^{(0)}$, $\mathbf{q}^{(0)}$, and $\mathbf{t}^{(0)}$ to zero
\STATE $\bar{\mathbf{f}}^{(0)} \leftarrow \mathbf{f}^{(0)}$
\REPEAT
\STATE  $\mathbf{w}^{(n+1)} = (\mathbf{w}^{(n)} + \sigma(\mathcal{H}\bar{\mathbf{f}}^{(n)} - \mathbf{g}^{[\mathcal{M}]}))/(1+\sigma)$
\STATE $\mathbf{p}^{\prime(n)} = \mathbf{p}^{(n)} + \sigma \nu_1 \mathcal{D}_x \bar{\mathbf{f}}^{(n)}$
\STATEx \hspace{0.25cm} $\mathbf{q}^{\prime(n)} = \mathbf{q}^{(n)} + \sigma \nu_2 \mathcal{D}_y \bar{\mathbf{f}}^{(n)}$
\STATE $\mathbf{p}^{(n+1)} = \mathbf{p}^{\prime(n)} - \sigma \frac{\mathbf{p}^{\prime(n)}}{|\mathbf{p}^{\prime(n)}|}\ell_1 {\rm ball}_{\nu_1 t_x} (\frac{\mathbf{p}^{\prime(n)}}{\sigma})$
\STATEx \hspace{0.25cm} $\mathbf{q}^{(n+1)} = \mathbf{q}^{\prime(n)} - \sigma \frac{\mathbf{q}^{\prime(n)}}{|\mathbf{q}^{\prime(n)}|}\ell_1 {\rm ball}_{\nu_2 t_y} (\frac{\mathbf{q}^{\prime(n)}}{\sigma})$
\STATE $\mathbf{t}^{(n+1)} = {\rm neg}({\mathbf{t}^{(n)} + \sigma \mu \bar{\mathbf{f}}^{(n)}})$
\STATE $\mathbf{f}^{(n+1)} = \mathbf{f}^{(n)}-\tau(\mathcal{H}^{\top}\mathbf{w}^{(n+1)}+\nu_1\mathcal{D}_x^{\top}{\mathbf{p}}^{(n+1)} +\nu_2\mathcal{D}_y^{\top}{\mathbf{q}}^{(n+1)} + \mu \mathbf{t}^{(n+1)})$
\STATE $\bar{\mathbf{f}}^{(n+1)} = 2 \mathbf{f}^{(n+1)}-\mathbf{f}^{(n)}$
\STATE $n \leftarrow n+1$
\UNTIL the convergence conditions are satisfied
\STATE OUTPUT: image $\mathbf{f}^{(n)}$
\end{algorithmic}
\end{algorithm}
In the pseudo-code, algorithm parameter $b$ is used for potentially improving the convergence rate of the DTV algorithm. In the studies below, we use $b\sim$ 1, 50, 100 and 200 to achieve reasonable convergence rates for angular ranges of $>\!180^\circ$, $120^\circ\!\sim\!150^\circ$, $60^\circ\!\sim\! 90^\circ$, and $14^\circ\!\sim\!30^\circ$, respectively. Matrix $\mathcal{K}$ has a transpose $\mathcal{K}^{\top} = (\mathcal{H}^{\top}, \nu_1 \mathcal{D}_x^{\top}, \nu_2 \mathcal{D}_y^{\top}, \mu \mathcal{I}$), in which the superscript ``$\top$'' indicates a transpose operation; $||\cdot||_2$ represents the largest singular value of a matrix; $\mathcal{I}$ is an identity matrix of size $N \times N$; vectors $\mathbf{w}^{(n)}$ is of size $M$, whereas vectors $\mathbf{p}^{\prime(n)}$, $\mathbf{q}^{\prime(n)}$, $\mathbf{p}^{(n)}$, $\mathbf{q}^{(n)}$, and $\mathbf{t}^{(n)}$ are of size $N$; operator ${\rm neg}(\cdot)$ enforces the non-positivity constraint; operator $\ell_1 {\rm ball}_{\beta}(\cdot)$ projects a vector onto the $\ell_1$-ball of scale $\beta$; $|\mathbf{q}^{\prime(n)}|$ depicts a vector of size $N$ with entry $j$ given by $(|\mathbf{q}^{\prime(n)}|)_j = |\mathbf{q}^{\prime(n)}_j|$; and $\mathbf{q}^{\prime(n)}_j$ indicates the $j$th entry of vector $\mathbf{q}^{\prime(n)}$.

\subsubsection{Numerical verification of the DTV algorithm}
While the DTV algorithm developed can solve the optimization program in Eq. \eqref{eq:opt}, the correctness of its computer implementation needs to be verified numerically, as it is the algorithm's computer implementation that is used in numerical studies and possible applications. We have performed a verification study, and include it in \ref{sec:Verify} again for avoiding distracting the presentation flow here. In the verification study, the DTV algorithm implemented is applied to reconstructing images from  data generated over a full-angular range of $2\pi$ from the breast phantom in Fig. \ref{fig:config}b. The rationale behind the verification-study design is that if the DTV algorithm is designed adequately and implemented correctly, it should yield  an image numerically identical to the truth image from which the full-angular-range data were generated. As evidenced quantitatively in \ref{sec:Verify},  the DTV algorithm is appropriately designed, and the correctness of its computer implementation is verified. Also, in all of the numerical studies in Secs. \ref{sec:Breast-result} and \ref{sec:bar-result} below, the results are obtained when the convergence conditions in Eqs. \eqref{eq:convergence_1} and \eqref{eq:convergence_2} in \ref{sec:convergence} are achieved up to the level of floating-precision.

\subsubsection{Reference reconstructions}
For reference, we also perform image reconstructions from limited-angular-range data by using the  FBP algorithm with a Hanning kernel and a 0.5 cut-off frequency, and an existing {TV algorithm} \citep{Sidky:06,Sidky:08,sidky2012convex,zhang2016artifact} that solves an image-TV constrained optimization program:
\begin{equation}\label{eq:opt-dl2tv}
\mathbf{f}^{\star} = \underset{\mathbf{f}}{\mathsf{argmin}} D_{\mathbf{g}}(\mathbf{f})  \quad
{\rm s.t.} \quad  || (|\nabla \mathbf{f}|) ||_1 \le t \,\,\, {\rm and} \,\,\, f_i \ge 0,
\end{equation}
where $|| (|\nabla \mathbf{f}|) ||_1$ denotes the image TV, and $t$ is the TV-constraint parameter. For differentiating from the DTV algorithm, we refer to the image TV as the {isotropic-TV} (ITV), and to the existing TV algorithm as the {\it ITV} algorithm in the work.

\section{Results: Reconstructions of the breast phantoms}\label{sec:Breast-result}
In this study, we consider image reconstruction of the breast phantom in Fig. \ref{fig:config}b, which is piece-wise constant,  and a blurred-breast phantom in Fig. \ref{fig:config}c, which is not piece-wise constant as it is obtained by convolving the breast phantom with a Gaussian convolver of a {  full-width-at-half-maximum (FWHM) $\sim\! 2$ image pixels.} Both breast phantoms are discretized on image arrays of $80\times 256$ {  square pixels of size 0.73 mm}. The directional and isotropic TVs of the breast and blurred-breast phantoms are computed and then used as the values of the constraint parameters in the DTV and ITV algorithms in the studies below. 

In the scanning configuration described in Fig. \ref{fig:config}a, we assume that the source-to-rotation distance (SRD) and source-to-detector distance (SDD) are 36 cm and 72 cm and that a linear detector composes $N_d=512$ bins of size $0.73$ mm, yielding a fan angle of $28.96^\circ$. Using the configuration, we generate data sets from the breast and blurred-breast phantoms over ten angular ranges, i.e., $\alpha=14^\circ$, $20^\circ$, $30^\circ$,  $60^\circ$, $90^\circ$,  $120^\circ$, $150^\circ$, $180^\circ$, $210^\circ$, and $360^\circ$, with angular interval of $ 1^\circ$ per view.

\begin{figure}
\centering
\includegraphics[angle=0,trim=0 0 0 0, clip,origin=c,width=1.0\textwidth]{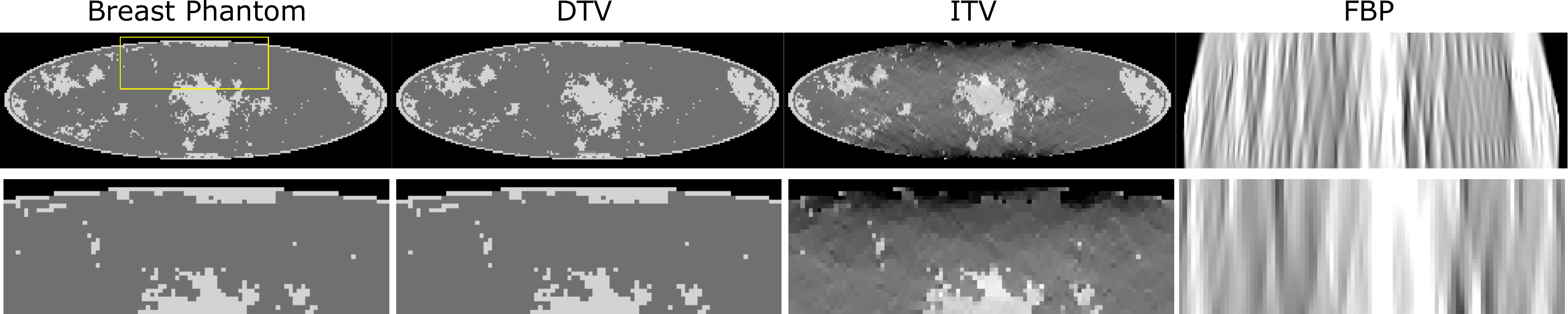}
\caption{Top row: images of the breast phantom and reconstructions obtained with the DTV, ITV, and FBP algorithms for the 20$^\circ$-angular range. Bottom row: zoomed-in views of DTV, ITV, and FBP reconstructions within a rectangular region of interest (ROI) of size 92$\times$32 indicated in the breast phantom in the top row. Display window: [0.15, 0.25] cm$^{-1}$.}
\label{fig:image-20-degs}
\end{figure}

\begin{figure}[h]
\centering
\includegraphics[angle=0,trim=0 0 0 0, clip,origin=c,width=1.\textwidth]{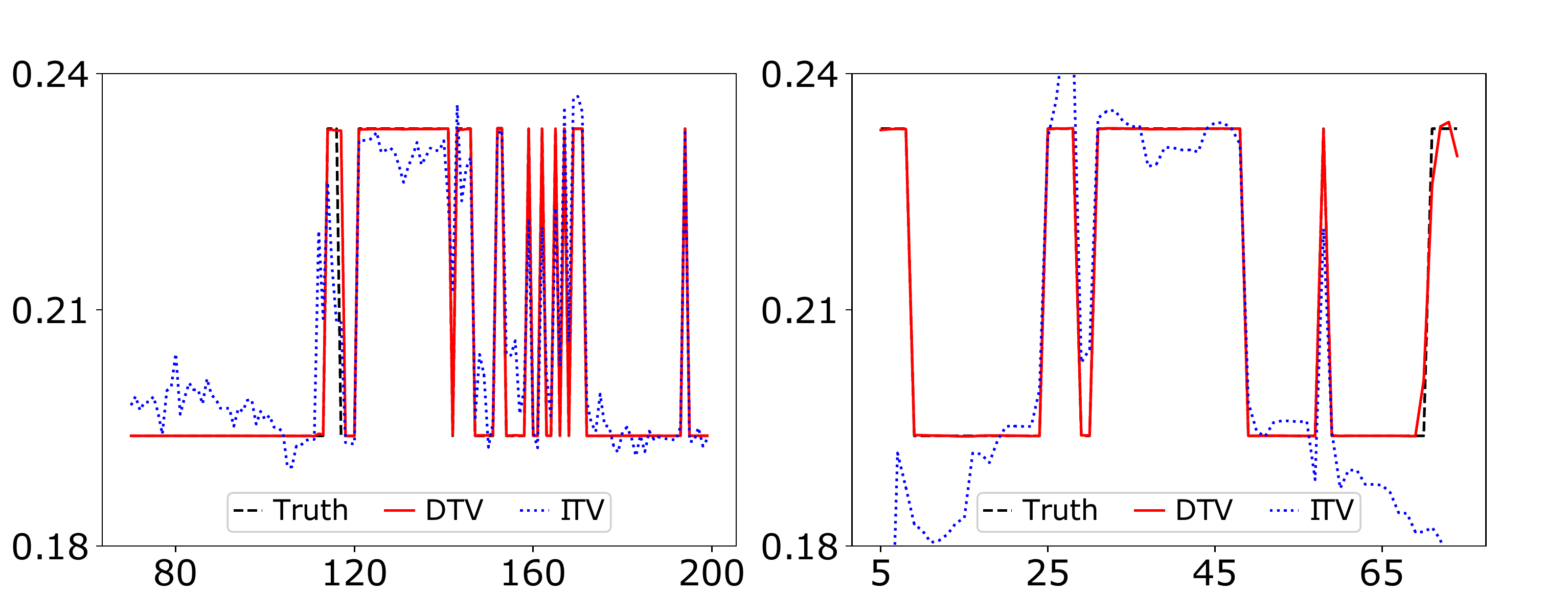}
\caption{Reconstruction profiles over the horizontal (left) and vertical (right) lines, depicted in the breast phantom in Fig. \ref{fig:config}b, obtained with the DTV (solid) and ITV  (dotted) algorithms from noiseless data generated over the 20$^\circ$-angular range. It can be observed that the DTV profiles coincide virtually completely with the corresponding truth profiles (dashed) of the breast phantom.}
\label{fig:breast-20-lineprofiles}
\end{figure}

\begin{figure}
\centering
\includegraphics[angle=0,trim=0 0 0 0, clip,origin=c,width=1.0\textwidth]{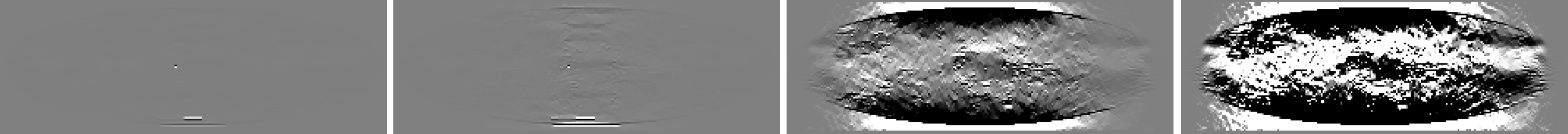}
\caption{Differences between truth and reconstructed images of the breast phantom with DTV (columns 1 \& 2) and ITV (columns 3 \& 4) algorithms from data over an angular range of $20^\circ$. Display window: [-$10^{-2}$, $10^{-2}$] cm$^{-1}$ for columns 1 \& 3, [-$10^{-3}$, $10^{-3}$] cm$^{-1}$ for columns 2 \& 4.}
\label{fig:breast-diff-20d}
\end{figure}

\subsection{Reconstruction of the breast phantom from $20^\circ$-data}\label{para:breast_20_degs}
We first apply the DTV algorithm, along with the FBP and ITV algorithms, to reconstructing images of the breast phantom from its data over an angular range of $20^\circ$, and display them in Fig. \ref{fig:image-20-degs}. It can be observed that the DTV algorithm minimizes artifacts observed in reconstructions of the ITV and FBP algorithms. In X-ray tomography such as DBT applications, it can be difficult to visualize appropriately, with a single display window, an image  especially within a transverse plane (i.e., within the $x$-$y$ plane depicted in Fig. \ref{fig:config}a) because the image obtained with existing algorithms \citep{rose2019imaging} contains significant artifacts in the form of intensity distortion. In  Fig. \ref{fig:breast-20-lineprofiles}, the DTV- and ITV-reconstruction profiles over the two lines depicted in the breast phantom in Fig.  \ref{fig:config}b confirm
 that the DTV reconstruction is more accurate than the ITV reconstruction as compared to the breast phantom. No profile results of the FBP reconstructions are shown because they are beyond the display range. For the same reason, we show no FBP results in additional studies on the breast phantoms below. Moreover, we show in Fig. \ref{fig:breast-diff-20d} differences between the breast-phantom image and reconstructions by use of DTV and ITV algorithms, respectively, in two display windows for further demonstrating that the DTV reconstruction is more accurate.

\begin{figure}
\centering
\includegraphics[angle=0,trim=0 0 0 0, clip,origin=c,width=1.0\textwidth]{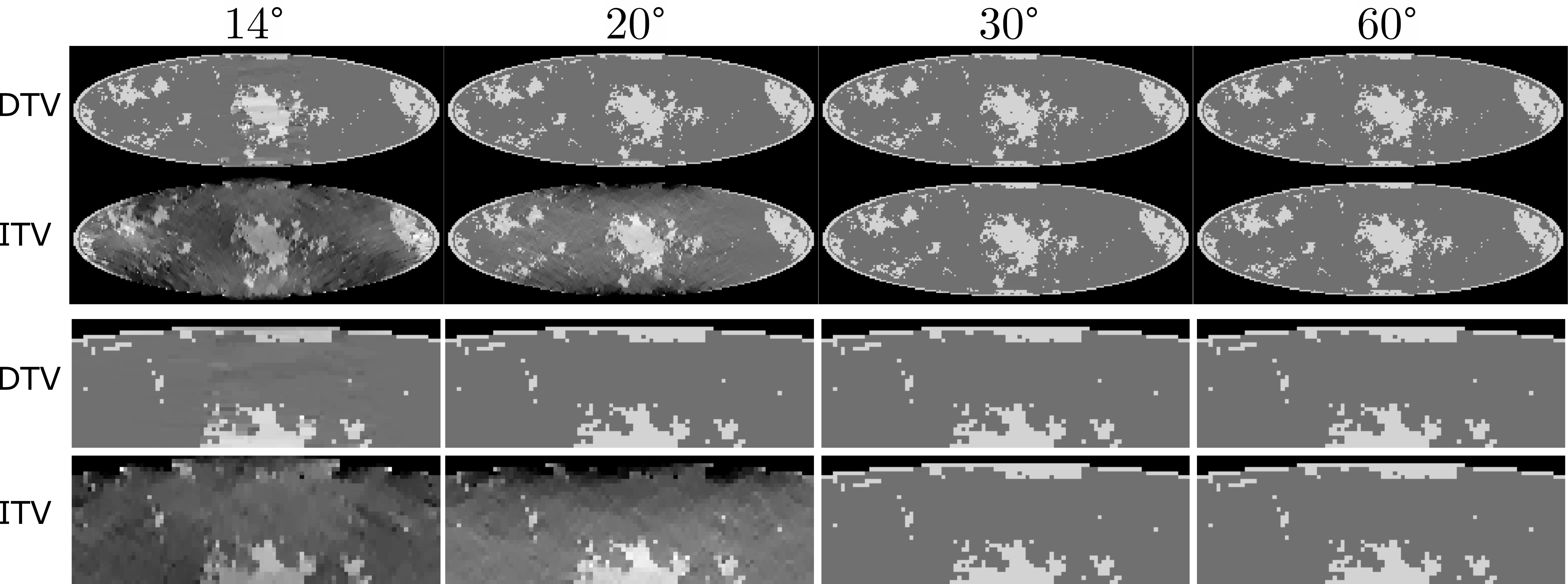}
\caption{Reconstructions of the breast phantom obtained with the DTV (row 1) and ITV  (row 2) algorithms over angular ranges of 14$^\circ$ (column 1), 20$^\circ$ (column 2), 30$^\circ$ (column 3), and 60$^\circ$ (column 4). Zoomed-in views of DTV (row 3) and ITV (row 4) images within a ROI of size 92$\times$32 indicated in the breast phantom in the top row of Fig. \ref{fig:image-20-degs}. Display window: [0.15, 0.25] cm$^{-1}$.}
\label{fig:breast-angular-NoNoise}
\end{figure}
\begin{figure}
\centering
\includegraphics[angle=0,trim=0 0 20 0, clip,origin=c,width=1.0\textwidth]{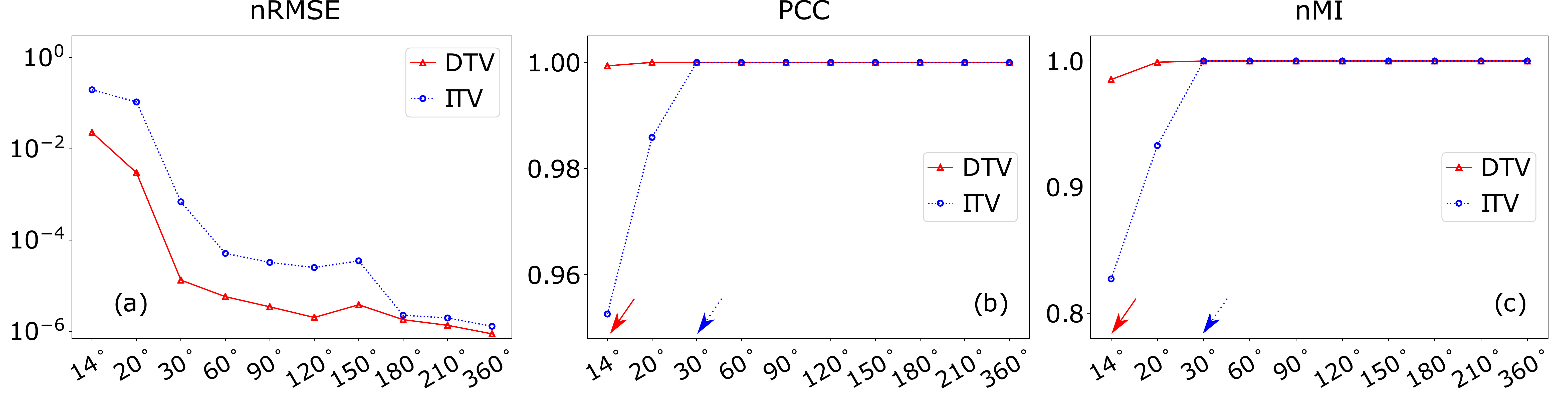}
\caption{Metrics nRMSE (a), PCC (b), and nMI (c) of the breast-phantom reconstructions obtained with the  DTV (triangle) and ITV (circle) algorithms, as functions of angular range $\alpha$.
The two arrows in (b) and (c) indicate empirical minimal-angular-ranges $\alpha_{\rm min}\!\sim\! 14^\circ$ and $\sim\! 30^\circ$ for the DTV and ITV algorithms, respectively.}
\label{fig:breast-angular-NoNoise-rmse}
\end{figure}

\subsection{Reconstructions of the breast phantom from additional limited-angular-range data}\label{sec:breast_angular_range}
In Fig. \ref{fig:breast-angular-NoNoise}, we show images, and their ROI images, reconstructed from data over angular ranges, i.e., $\alpha=14^\circ$, $20^\circ$, $30^\circ$, and $60^\circ$, respectively, obtained with the DTV and ITV algorithms.  As expected, reconstruction artifacts diminish as the angular range increases, and the DTV and ITV reconstructions appear to be visually comparable for $\alpha \ge 30^\circ$. However,  for $14^\circ\le \alpha < 30^\circ$, the latter contains visible artifacts, which are removed largely in the former. 

We also compute metrics normalized root-mean-square-error (nRMSE), Pearson correlation coefficient (PCC), and normalized mutual information (nMI) \citep{pearson1895notes,Viergever:2003,Bian-PMB:2010} in Eqs. \eqref{eq:imdist}-\eqref{eq:nMI} in \ref{sec:Verify} to assess quantitatively the reconstructions in which the reference image is the truth image, i.e., the breast phantom. When the nRMSE is near zero, it provides a meaningful global measure of reconstruction accuracy, but it is also known to be a poor measure for visualization correlation between the reconstruction and reference images when it is not approaching zero. Conversely, metrics PCC and nMI may provide a more direct, meaningful measure of visualization correlation between reconstruction and reference images, as shown below. 

In Fig. \ref{fig:breast-angular-NoNoise-rmse}, we display metrics nRMSE, PCC, and nMI obtained as functions of angular range $\alpha$. It can be observed in Fig. \ref{fig:breast-angular-NoNoise-rmse}a that the nRMSE of the DTV reconstruction is lower than that of the ITV reconstruction for all of the angular ranges considered, and that the nRMSE differences between the two reconstructions diminish, as expected, for large angular ranges (e.g., $\ge 180^\circ$).  
Metrics PCC and nMI in Figs. \ref{fig:breast-angular-NoNoise-rmse}b and \ref{fig:breast-angular-NoNoise-rmse}c provide measures of the visual correlation between the reconstruction and reference images. In particular, when PCC and nMI approach 1, the reconstruction and reference images appear visually indistinguishable. As the PCC and nMI results  in Figs. \ref{fig:breast-angular-NoNoise-rmse}b and \ref{fig:breast-angular-NoNoise-rmse}c reveal, the DTV reconstructions visually correlate better to the reference image at limited-angular ranges ($< 30^\circ$) than the ITV reconstructions. Inspecting the nRMSE, PCC, and nMI results, we can obtain empirical minimal-angular ranges, highlighted by two arrows in Figs. \ref{fig:breast-angular-NoNoise-rmse}b and \ref{fig:breast-angular-NoNoise-rmse}c, sufficient for numerically and visually accurate reconstructions by use of the DTV and ITV  algorithms, which are $\alpha_{\rm min}\!\sim\! 14^\circ$ and $\sim\! 30^\circ$, respectively, for the breast phantom.

\subsection{Reconstruction of the blurred-breast phantom from limited-angular-range data}\label{sec:blurred_breast_angular_range}

\begin{figure}[t]
\centering
\includegraphics[angle=0,trim=0 0 0 0, clip,origin=c,width=1.0\textwidth]{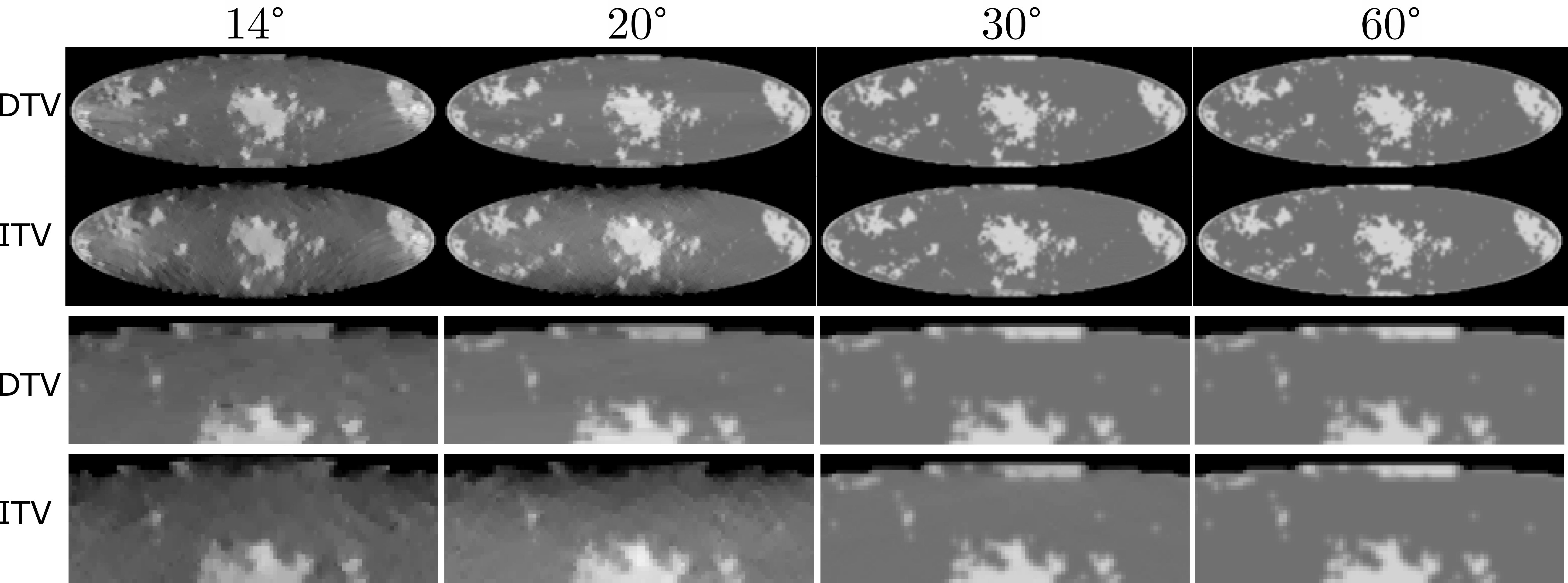}
\caption{Reconstructions of the blurred-breast phantom obtained with the DTV (row 1) and ITV  (row 2) algorithms over angular ranges of 14$^\circ$ (column 1), 20$^\circ$ (column 2), 30$^\circ$ (column 3), and 60$^\circ$ (column 4). Zoomed-in views of DTV (row 3) and ITV (row 4) images within a ROI of size 92$\times$32 indicated in the breast phantom in the top row of Fig. \ref{fig:image-20-degs}. Display window: [0.15, 0.25] cm$^{-1}$.}
\label{fig:breast-angular-blur}
\end{figure}

We display in Fig. \ref{fig:breast-angular-blur} reconstructions and their ROIs of the blurred-breast phantom, which is not piece-wise constant, from data over angular ranges $\alpha=14^\circ$, $20^\circ$, $30^\circ$, and $60^\circ$, respectively, obtained with the DTV and ITV algorithms.  It can be observed again that reconstruction artifacts diminish as the angular range increases, that the DTV and ITV  reconstructions appear to be visually comparable for $\alpha \ge 60^\circ$, and that the former can considerably reduce the artifacts visible in reconstructions of the latter for $\alpha<60^\circ$.  

\begin{figure}
\centering
\includegraphics[angle=0,trim=0 0 20 0, clip,origin=c,width=1.0\textwidth]{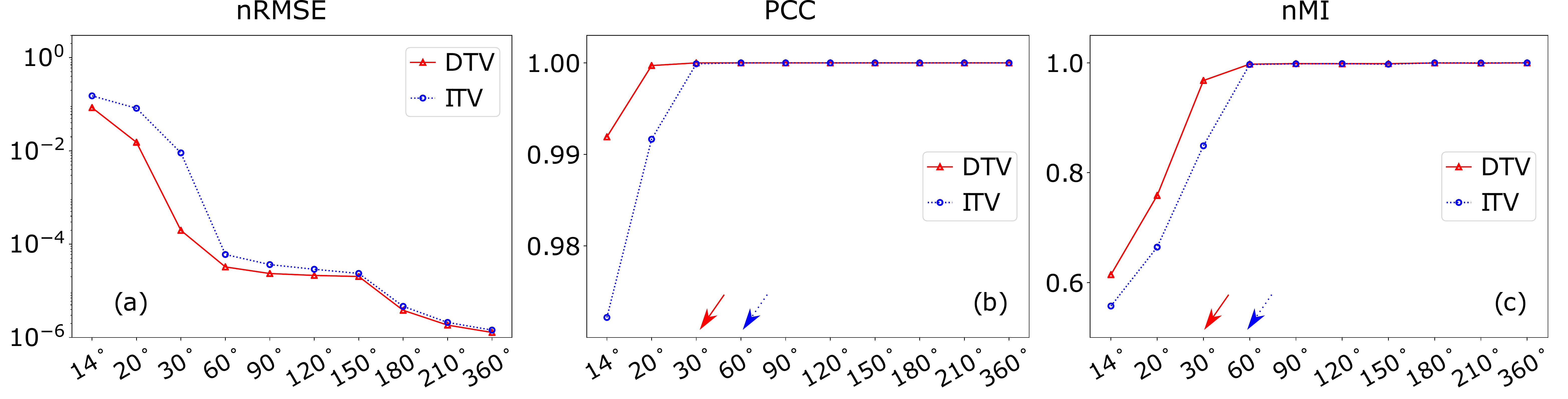}
\caption{Metrics nRMSE (a), PCC (b), and nMI (c) of the blurred-breast-phantom obtained with the  DTV (triangle) and ITV (circle) algorithms, as functions of angular range $\alpha$.
The two arrows in (b) and (c) indicate empirical minimal-angular-ranges $\alpha_{\rm min}\!\sim\! 30^\circ$ and $\sim\! 60^\circ$ for the DTV and ITV algorithms, respectively.}
\label{fig:blurred-breast-angular-NoNoise-rmse}
\end{figure} 


In Fig. \ref{fig:blurred-breast-angular-NoNoise-rmse}, we display metrics nRMSE, PCC, and nMI calculated as functions of angular range $\alpha$ for the blurred-breast phantom. Observations similar to those for the breast phantom above can be  made for the blurred-breast phantom. However, as shown in Fig. \ref{fig:blurred-breast-angular-NoNoise-rmse}a, the nRMSEs of the blurred-breast phantom are higher than their counterparts for the breast phantom in Fig. \ref{fig:breast-angular-NoNoise-rmse}a. This result can be understood as follows: from the compressive sensing perspective, the higher the numbers of non-zeros in the gradient-magnitude images (GMIs,) the more difficult to invert accurately the linear data model considered.  The number of non-zeros in the directional-GMIs of the blurred-breast phantom is substantially higher than that of non-zeros in the directional-GMIs of the breast phantom, and thus the nRMSEs obtained with the DTV algorithm for the blurred-breast phantom are higher than that for the breast phantom for a given limited-angular range.  

On the other hand, it can be observed that the nRMSEs of the ITV reconstructions for the breast and blurred-breast phantoms appear largely comparable despite the fact that the numbers of non-zeros in their isotropic-GMIs are significantly different. This is because for the angular range $\alpha <180^\circ$, the data-model ill-conditionedness is the dominant factor, instead of  the number of non-zeros in an isotropic-GMI, that impacts the performance of the ITV algorithm.  Therefore, for a given angular range $\alpha <180^\circ$, the nRMSE results obtained with the ITV algorithm are comparable for the breast and blurred-breast phantoms.

Again, in addition to nRMSE, we also plot PCC and nMI in Fig. \ref{fig:blurred-breast-angular-NoNoise-rmse}b and \ref{fig:blurred-breast-angular-NoNoise-rmse}c. Based upon the quantitative results, we can obtain empirical minimal-angular ranges, highlighted by two arrows in Fig. \ref{fig:blurred-breast-angular-NoNoise-rmse}b and \ref{fig:blurred-breast-angular-NoNoise-rmse}c, sufficient for numerically and visually accurate reconstructions with the DTV and ITV algorithms, which are $\alpha_{\rm min}\!\sim\! 30^\circ$ and $\sim\! 60^\circ$, respectively, for the blurred-breast phantom. 

\begin{figure}[h!]
\centering
\includegraphics[angle=0,trim=0 0 0 0, clip,origin=c,width=1.\textwidth]{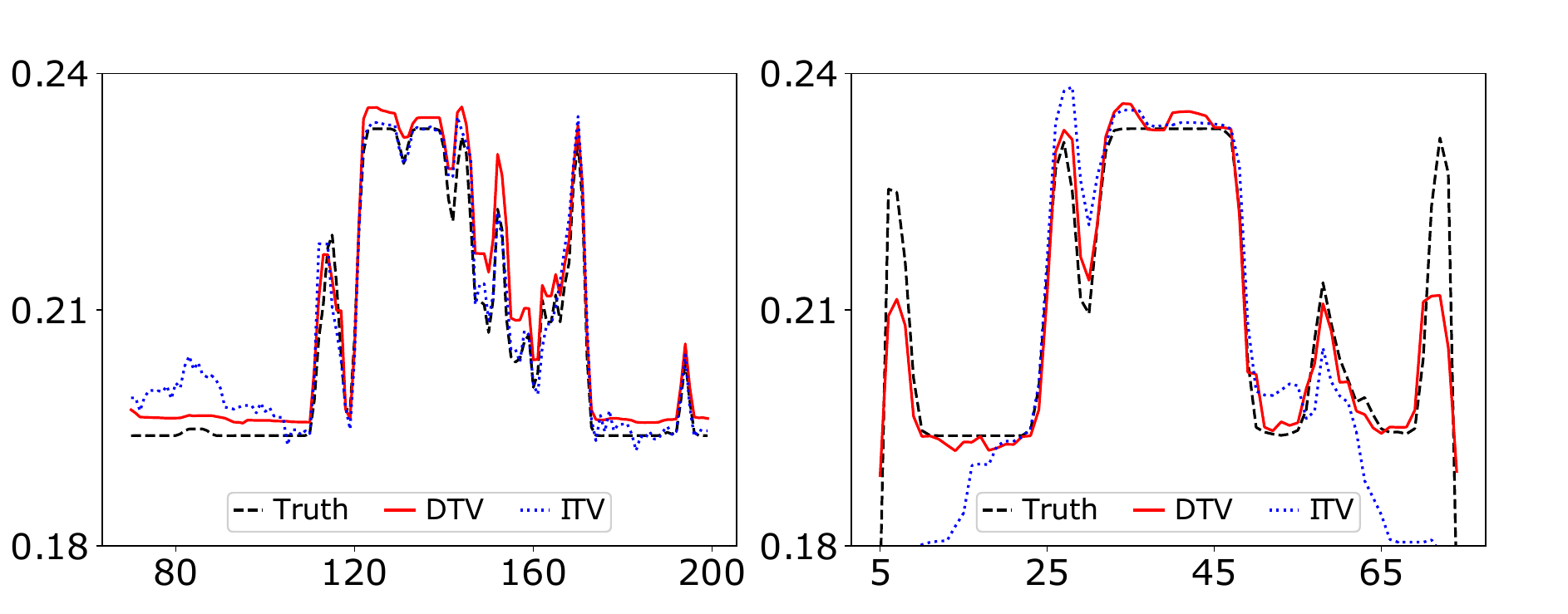}
\caption{{  Reconstruction profiles over the horizontal (left) and vertical (right) lines, depicted in the blurred-breast phantom in Fig. \ref{fig:config}c, obtained with the DTV (solid) and ITV  (dotted) algorithms from noiseless data generated over the 20$^\circ$-angular range. It can be observed that the DTV profiles agree with the corresponding truth profiles (dashed) of the burred-breast phantom more closely than the ITV profiles.}}
\label{fig:blur_breast-20-lineprofiles}
\end{figure}
{  For further revealing reconstruction details, we display in Fig. \ref{fig:blur_breast-20-lineprofiles} the profiles of DTV and ITV reconstructions  of the blurred-breast phantom (in column 2 of Fig. \ref{fig:breast-angular-blur}) from $20^\circ$-data over the two lines that overlay the blurred-breast phantom in Fig. \ref{fig:config}c. We observe that the profiles in the DTV reconstruction follow the truth profiles better than do those in the ITV reconstruction. Moreover, the DTV profiles from data over $\alpha_{\rm min}\ge 30^\circ$ coincide virtually completely with their corresponding truth profiles of the blurred-breast phantom.}

\begin{figure}[h]
\centering
\includegraphics[angle=0,trim=0 0 0 0, clip,origin=c,width=1.0\textwidth]{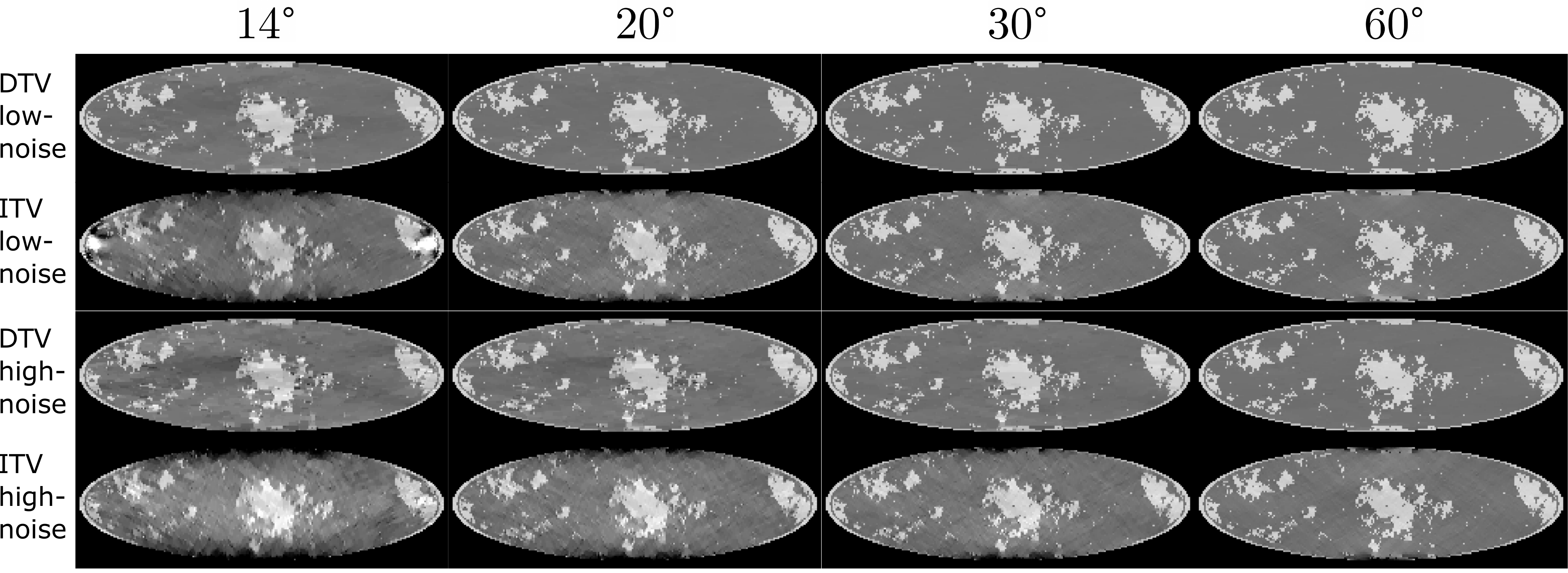}
\caption{{Images of the breast phantom reconstructed by use of the DTV (rows 1 $\&$ 3) and ITV  (rows 2 $\&$ 4) algorithms from data of low- (rows 1 $\&$ 2) and high- (rows 3 $\&$ 4) noise levels for angular ranges of 14$^\circ$ (column 1), 20$^\circ$ (column 2), 30$^\circ$ (column 3), and 60$^\circ$ (column 4). Display window: [0.15, 0.25] cm$^{-1}$.}}
\label{fig:breast-angular-noisy}
\end{figure}
\begin{figure}
\centering
\includegraphics[angle=0,trim=0 0 0 0, clip,origin=c,width=1.0\textwidth]{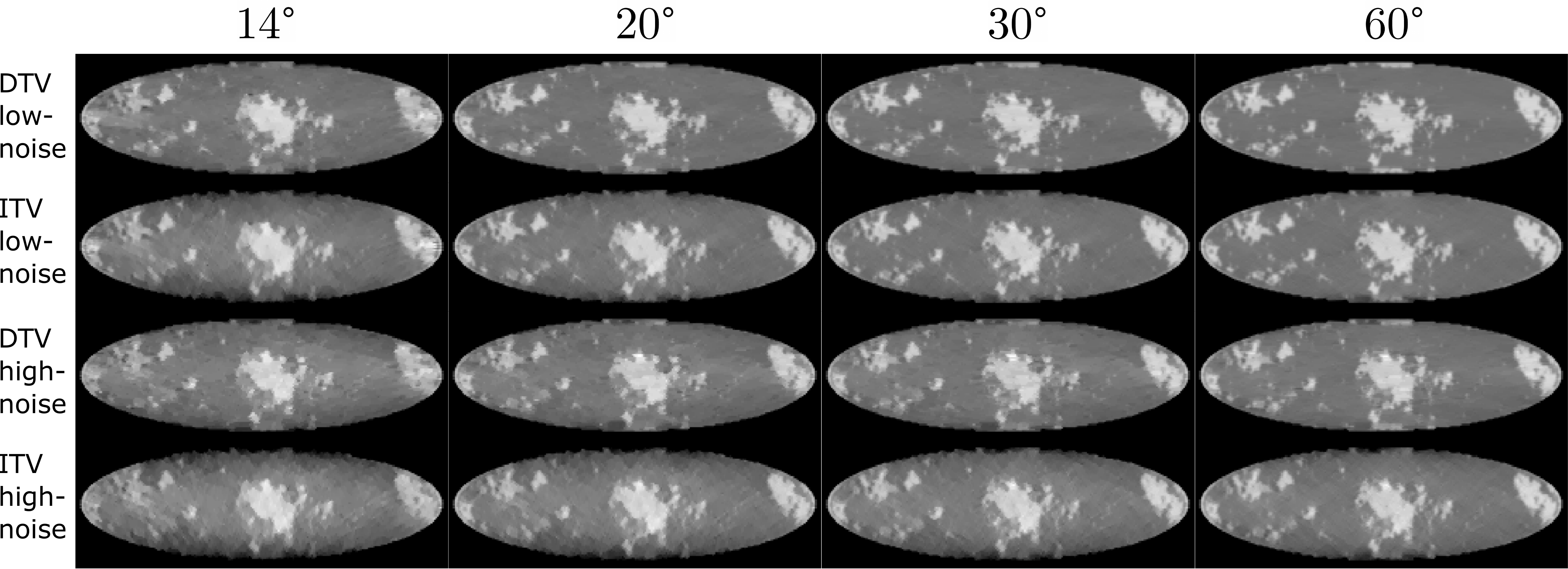}
\caption{{Images of the blurred-breast phantom reconstructed by use of the DTV (rows 1 $\&$ 3) and ITV (rows 2 $\&$ 4) algorithms from  data of low- (rows 1 $\&$ 2) and high- (rows 3 $\&$ 4) noise levels for angular ranges of 14$^\circ$ (column 1), 20$^\circ$ (column 2), 30$^\circ$ (column 3), and 60$^\circ$ (column 4). Display window: [0.15, 0.25] cm$^{-1}$.}}
\label{fig:blurred-breast-angular-noisy}
\end{figure}

\subsection{Reconstruction of the breast phantoms from noisy limited-angular-range data } \label{sec:breast_noise}
We perform a preliminary investigation of image reconstruction from noisy, limited-angular-range data. Using each of the noiseless-data sets from the breast or blurred-breast phantom described in Sec. \ref{sec:breast_angular_range} as the mean, we generate noisy-data sets of multiple Poisson-noise levels, and then reconstruct images from the noisy-data sets. The study results in Fig. \ref{fig:breast-angular-noisy} show reconstructions only from data sets of low- and high-noise levels with signal-to-noise-ratios (SNRs) of $\sim\!10^4$ and $\sim\!3.2\times10^3$, resulting in $\sim\!10^8$ and $\sim\!10^7$ emitting photons for each ray, for four limited-angular ranges $\alpha = 14^\circ$, $20^\circ$, $30^\circ$, and $60^\circ$, respectively. The results of other angular ranges and noise levels are not included because observations similar to those presented below can be made. 

It can be observed in Fig. \ref{fig:breast-angular-noisy} that reconstructions obtained with the DTV algorithm in general show reduced artifacts comparing to those obtained with the ITV algorithm. As expected, the reconstruction quality diminishes as the angular range decreases from 60$^\circ$ to 14$^\circ$, and the lower the level of data noise, the less artifacts are observed in the reconstructions.  Notably, reconstruction of the DTV algorithm is more robust than that of the ITV algorithm as the former contains visually less severe artifacts than does the latter. We have also computed metrics nRMSE, PCC, and nMI of these reconstructions, which are not shown here, and observe that, in terms of the metrics,  the DTV reconstruction yields reconstructions quantitatively and visually better than that of the ITV reconstructions. 

In Fig. \ref{fig:blurred-breast-angular-noisy}, we also display images of the blurred-breast phantom reconstructed from data of low- and high-noise levels, respectively, by using the DTV and ITV algorithms.  Observations similar to those for the breast phantom can be made for the blurred-breast phantom.

\begin{figure}[h]
\centering
\includegraphics[angle=0,trim=0 0 0 0, clip,origin=c,width=1.0\textwidth]{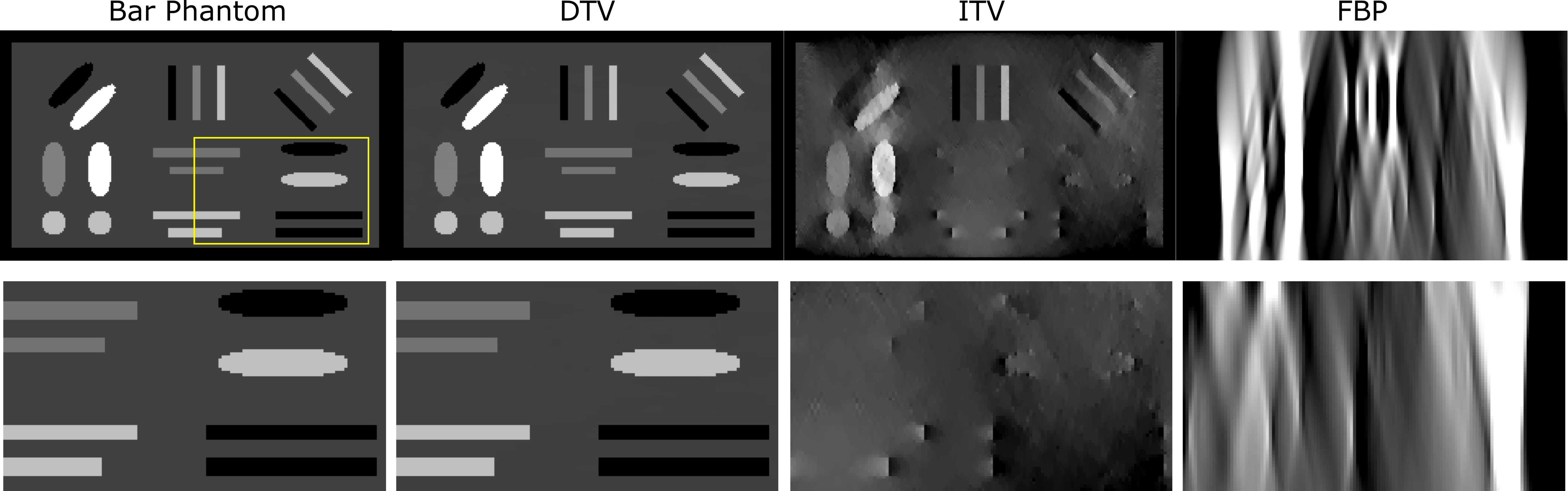}
\caption{Top row: images of the bar phantom and reconstructions obtained with the DTV, ITV, and FBP algorithms for the 20$^\circ$-angular range. Bottom row: zoomed-in views of DTV, ITV, and FBP reconstructions within a rectangular region of interest (ROI) of size 128$\times$70 indicated in the bar phantom in the top row. Display window: [0.1, 0.5] cm$^{-1}$.}
\label{fig:bar-image-20-degs}
\end{figure}
\begin{figure}[h]
\centering
\includegraphics[angle=0,trim=0 0 0 0, clip,origin=c,width=1.\textwidth]{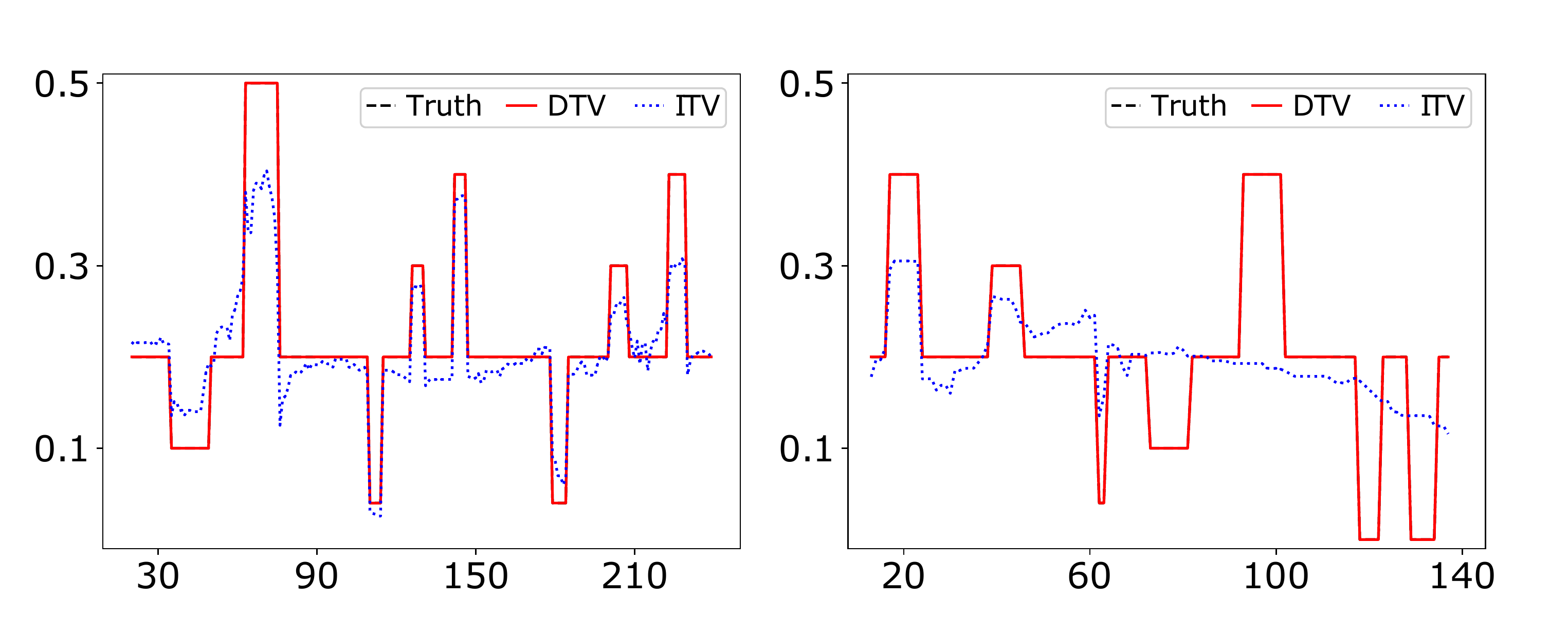}
\caption{Reconstruction profiles over the horizontal (left) and vertical (right) lines, depicted in the bar phantom in Fig. \ref{fig:config}d, obtained with the DTV (solid) and ITV  (dotted) algorithms from noiseless data generated over the 20$^\circ$-angular range. It can be observed that the DTV profiles coincide virtually completely with  the corresponding truth profiles (dashed) of the bar phantom.}
\label{fig:bar-profiles-20-degs}
\end{figure}

\begin{figure}
\centering
\includegraphics[angle=0,trim=0 0 0 0, clip,origin=c,width=1.0\textwidth]{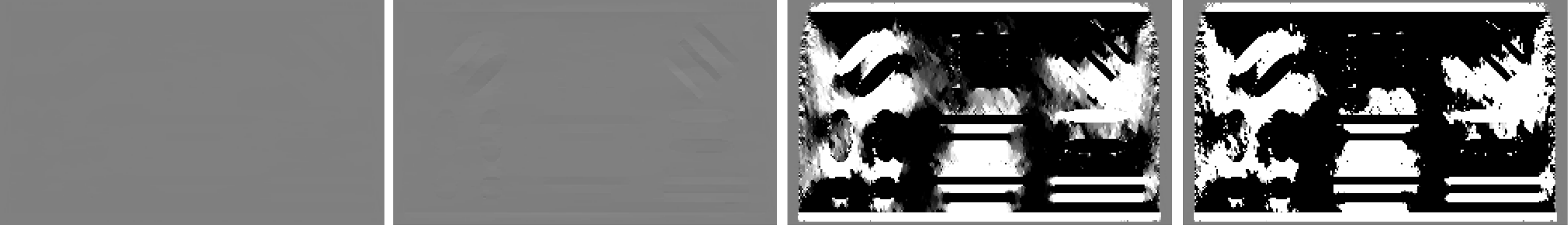}
\caption{Differences between truth and reconstructed images of the bar phantom with DTV (columns 1 \& 2) and ITV (columns 3 \& 4) algorithms from data over an angular range of $20^\circ$. Display window: [-$10^{-2}$, $10^{-2}$] cm$^{-1}$ for columns 1 \& 3, [-$10^{-3}$, $10^{-3}$] cm$^{-1}$ for columns 2 \& 4.}
\label{fig:bar-diff-20d}
\end{figure}

\section{Results: Reconstructions of the bar phantoms}\label{sec:bar-result}

In industrial applications of X-ray tomographic imaging, objects of interest often contain high-contrast, strip-shape structures. In the study below, we consider a bar phantom in Fig. \ref{fig:config}d, which is piece-wise constant, and a blurred-bar phantom in Fig. \ref{fig:config}e, which is not piece-wise constant as it is obtained by convolving the bar phantom with a Gaussian convolver of {  a FWHM of $\sim\! 2$ image pixels.} Both bar phantoms are discretized on image arrays of $150\times 256$ {  square pixels of size 1.38 mm}.  The directional and isotropic TVs of the bar and blurred-bar phantoms are computed and then used as the values of the constraint parameters on the DTV and ITVs in the bar-phantom studies below. 


Using the scanning configuration in Fig. \ref{fig:config}a, now with SRD=100 cm, SDD=150 cm, a linear detector containing $N_d=512$ bins of size $1.38$ mm, and forming a fan angle of $22.14^\circ$. We generate data sets from the bar and  blurred-bar phantoms over ten angular ranges, i.e., $\alpha=14^\circ$, $20^\circ$, $30^\circ$,  $60^\circ$, $90^\circ$,  $120^\circ$, $150^\circ$, $180^\circ$, $210^\circ$, and $360^\circ$, with angular interval of $1^\circ$ per view.

\subsection{Reconstruction of the bar phantom from $20^\circ$ data}\label{para:bar_20_degs}

We show in Figs. \ref{fig:bar-image-20-degs} and \ref{fig:bar-profiles-20-degs} the images of the bar phantom and its profiles reconstructed with the DTV, ITV, and FBP algorithms  from its data generated over an angular range of $20^\circ$.  It can be observed in Fig. \ref{fig:bar-image-20-degs} that the DTV algorithm reduces considerably artifacts observed in images of the ITV and FBP algorithms, as also corroborated by the reconstruction profiles in Fig. \ref{fig:bar-profiles-20-degs} over the two lines depicted in the bar phantom. The profiles of the FBP reconstruction are not shown because they are beyond the display range, and we show no additional FBP results in additional studies on the bar phantoms below. Differences between the bar-phantom image and reconstructions by use of DTV and ITV algorithms, respectively, are shown in Fig. \ref{fig:bar-diff-20d} in two display windows for further demonstrating that the DTV reconstruction is more accurate.

{\begin{figure}[h!]
\centering
\includegraphics[angle=0,trim=0 0 0 0, clip,origin=c,width=1.0\textwidth]{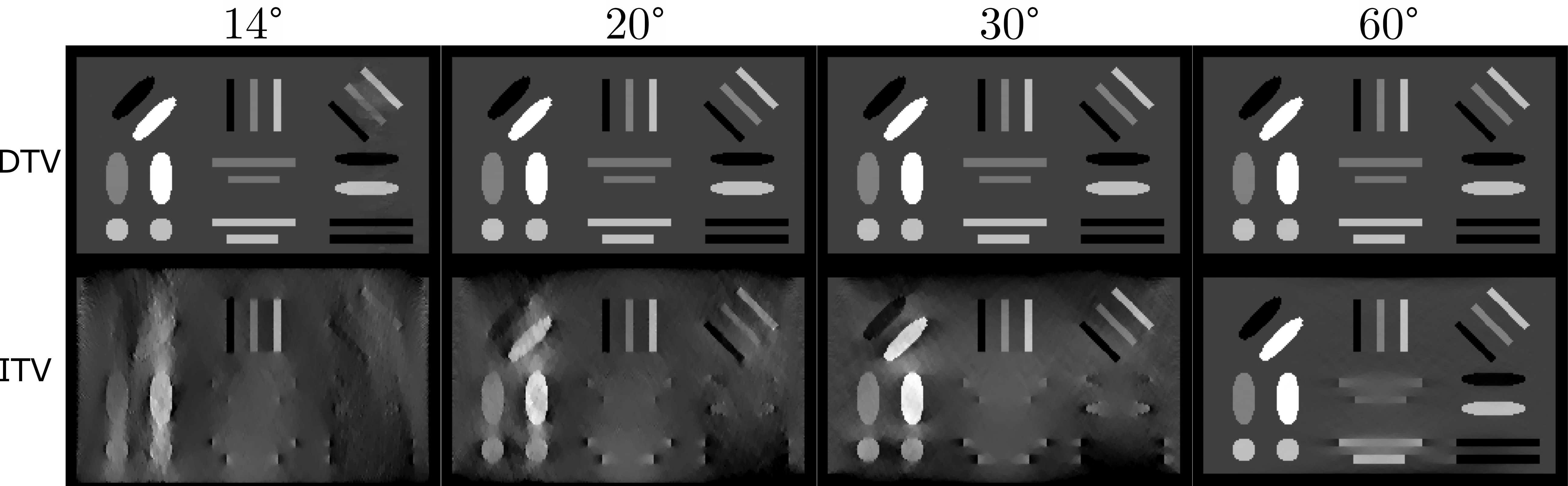}
\caption{{Images of the bar phantom reconstructed by use of the DTV (top row) and ITV (bottom row) algorithms over angular ranges of 14$^\circ$ (column 1), 20$^\circ$ (column 2), 30$^\circ$ (column 3), and 60$^\circ$ (column 4). Display window: [0.1, 0.5] cm$^{-1}$.}}
\label{fig:bar-angular-NoNoise}
\end{figure}
\begin{figure}[h]
\centering
\includegraphics[angle=0,trim=0 0 20 0, clip,origin=c,width=1.0\textwidth]{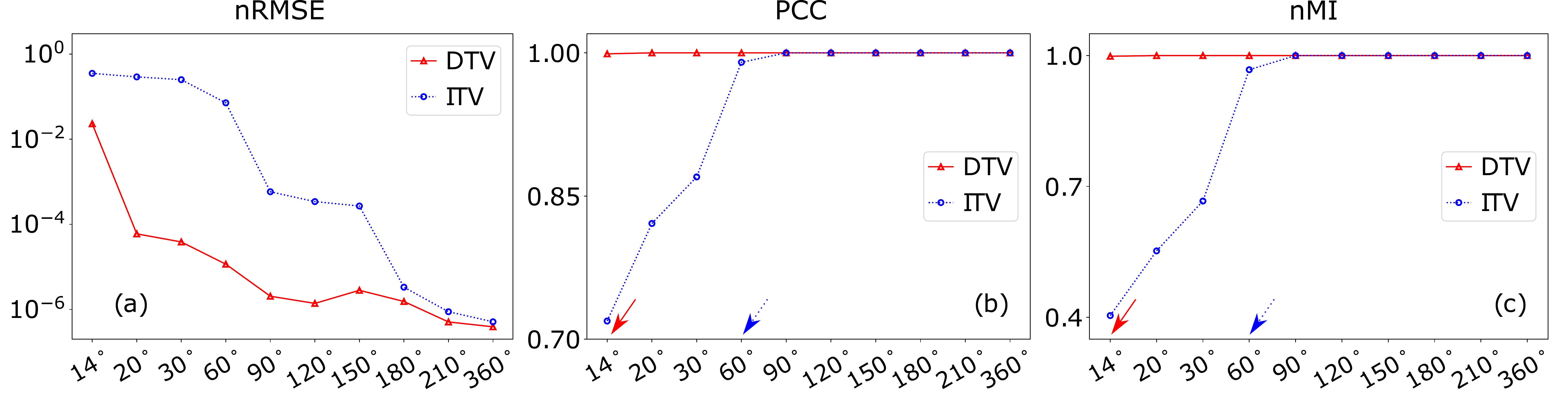}
\caption{Metrics nRMSE (a), PCC (b), and nMI (c) of the bar-phantom reconstruction obtained with the  DTV (triangle) and ITV (circle) algorithms, as functions of angular range $\alpha$. The two arrows in (b) and (c) indicate empirical minimal-angular-ranges $\alpha_{\rm min}\!\sim\! 14^\circ$ and $\sim\! 60^\circ$ for the DTV and ITV algorithms, respectively.}
\label{fig:bar-angular-NoNoise-rmse}
\end{figure}
}

\subsection{Reconstructions of the bar phantom from additional limited-angular-range data}\label{sec:bar_angular_range}

In the top and bottom rows of Fig. \ref{fig:bar-angular-NoNoise}, we show images reconstructed from data over  angular ranges $\alpha=14^\circ$, $20^\circ$, $30^\circ$, and $60^\circ$, respectively, obtained with the DTV and ITV algorithms.  Reconstruction artifacts diminish as the angular range increases, as expected, and the DTV and ITV reconstructions appear visually comparable for $\alpha > 60^\circ$. However,  for $ 14^\circ\le \alpha\le 60^\circ$, the former shows little artifacts observable in the latter. 

We display in Fig. \ref{fig:bar-angular-NoNoise-rmse} metrics nRMSE, PCC, and nMI of the bar phantom as functions of angular range $\alpha$. The nRMSE results reveal quantitatively that while the DTV and ITV  algorithms yield reconstructions of comparable accuracy for $\alpha > 180^\circ$, the former reconstructs images more accurately than does the latter for $14^\circ\le \alpha \le 180^\circ$. It can also be observed that the nRMSE difference between the two algorithms decreases for large angular ranges (e.g., $\alpha \ge 150^\circ$,) as expected. 
As the PCC and nMI results  in Figs. \ref{fig:bar-angular-NoNoise-rmse}b and \ref{fig:bar-angular-NoNoise-rmse}c show,  the DTV reconstructions visually correlate better to the reference image at limited-angular ranges than the ITV reconstructions.
Basing upon the nRMSE, PCC, and nMI results, we can obtain empirical minimal-angular ranges, indicated by two arrows in Figs. \ref{fig:bar-angular-NoNoise-rmse}b and \ref{fig:bar-angular-NoNoise-rmse}c, sufficient for numerically and visually accurate reconstruction with the DTV and ITV  algorithms, which are $\alpha_{\rm min}\!\sim\! 14^\circ$ and $\sim\! 60^\circ$, respectively, for the bar phantom.

\subsection{Reconstruction of the blurred-bar phantom from limited-angular-range data}\label{sec:blurred_bar_angular_range}

\begin{figure}[h]
\centering
\includegraphics[angle=0,trim=0 0 0 0, clip,origin=c,width=1.0\textwidth]{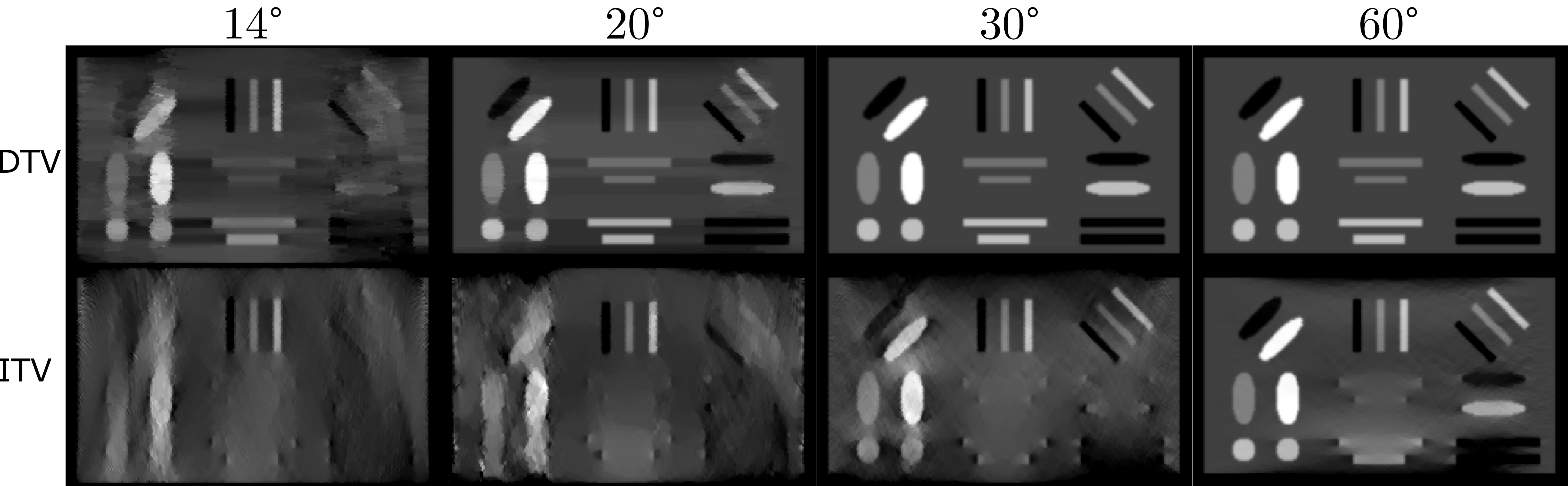}
\caption{{Images of the blurred-bar phantom reconstructed by use of the DTV (top row) and ITV  (bottom row) algorithms over angular ranges of 14$^\circ$ (column 1), 20$^\circ$ (column 2), 30$^\circ$ (column 3), and 60$^\circ$ (column 4). Display window: [0.1, 0.5] cm$^{-1}$.}}
\label{fig:bar-angular-blur}
\end{figure}

In the top and bottom rows of Fig. \ref{fig:bar-angular-blur}, we display reconstructions for $\alpha=14^\circ$, $20^\circ$, $30^\circ$, and  $60^\circ$ by use of the DTV and ITV algorithms.  Again, it can be observed that image artifacts diminish as the angular range increases, that the DTV and ITV algorithms appear to yield visually comparable reconstructions for $\alpha>150^\circ$, and that the former can considerably reduce the artifacts visible in reconstructions of the latter for $\alpha<150^\circ$. 

\begin{figure}[t!]
\centering
\includegraphics[angle=0,trim=0 0 20 0, clip,origin=c,width=1.0\textwidth]{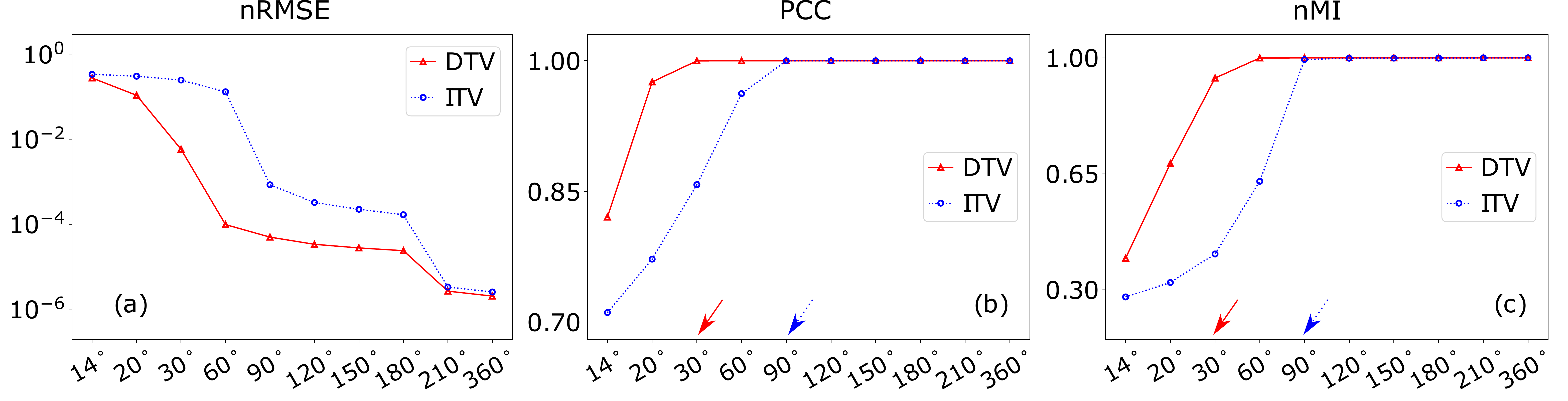}
\caption{Metrics nRMSE (a), PCC (b), and nMI (c) of the blurred bar-phantom reconstruction obtained with the  DTV (triangle) and ITV (circle) algorithms, as functions of angular range $\alpha$. The two arrows in (b) and (c) indicate empirical minimal-angular-ranges $\alpha_{\rm min}\!\sim\! 30^\circ$ and $\sim\! 90^\circ$ for the DTV and ITV algorithms, respectively.}
\label{fig:blurred-bar-angular-NoNoise-rmse}
\end{figure}

From the results of metrics nRMSE, PCC, and nMI in Fig. \ref{fig:blurred-bar-angular-NoNoise-rmse}, we can make observations similar to those for the bar phantom above. Additionally, as shown in Fig. \ref{fig:blurred-bar-angular-NoNoise-rmse}a, the nRMSEs of the blurred-bar phantom are higher than their counterparts for the bar phantom in Fig. \ref{fig:bar-angular-NoNoise-rmse}a. The reason for this increase in nRMSE for the blurred-phantom is the same as that illustrated for the blurred-breast phantom in Sec. \ref{sec:blurred_breast_angular_range} above. Moreover, it can be observed that the nRMSEs of the ITV reconstructions for the bar and blurred-bar phantoms appear largely comparable despite the fact that the numbers of non-zeros in their isotropic-GMIs  are significantly different, again, for the reason explained for the case of the blurred-breast phantom in Sec. \ref{sec:blurred_breast_angular_range} above. Using the  nRMSE, PCC, and nMI results in Fig. \ref{fig:blurred-bar-angular-NoNoise-rmse},
we obtain empirical minimal-angular ranges, depicted by two arrows, sufficient for numerically and visually accurate reconstruction with the DTV and ITV algorithms, which are $\alpha_{\rm min}\!\sim\! 30^\circ$ and $\sim\! 90^\circ$, respectively, for the blurred-bar phantom. 

\begin{figure}[h]
\centering
\includegraphics[angle=0,trim=0 0 0 0, clip,origin=c,width=1.\textwidth]{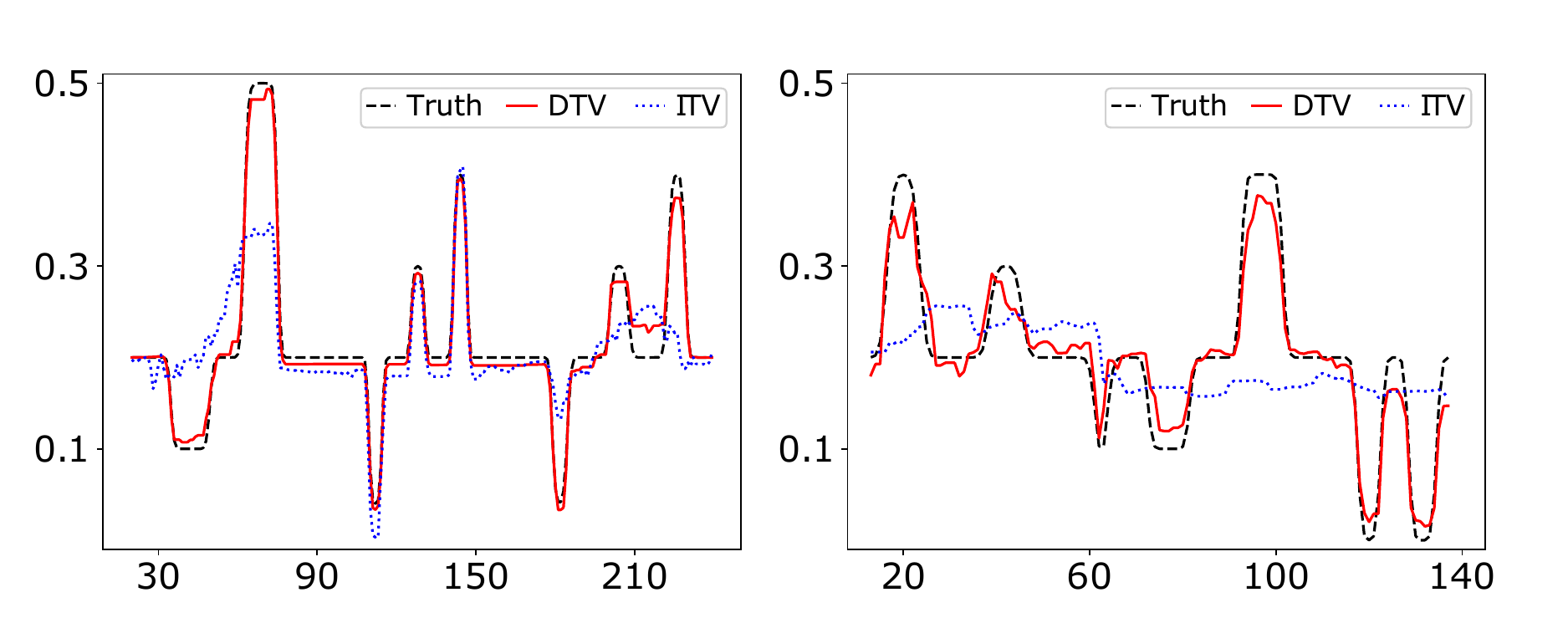}
\caption{{  Reconstruction profiles over the horizontal (left) and vertical (right) lines, depicted in the blurred-bar phantom in Fig. \ref{fig:config}e, obtained with the DTV (solid) and ITV  (dotted) algorithms from noiseless data generated over the 20$^\circ$-angular range. It can be observed that the DTV profiles agree with  the corresponding truth profiles (dashed) of the burred-bar phantom more closely than the ITV profiles.}}
\label{fig:blur_bar-profiles-20-degs}
\end{figure}

{  For further revealing reconstruction details, we display in Fig. \ref{fig:blur_bar-profiles-20-degs} the profiles of DTV and ITV reconstructions of the blurred-bar phantom (in column 2 of Fig. \ref{fig:bar-angular-blur}) from $20^\circ$-data over the two lines that overlay the blurred-bar phantom in Fig. \ref{fig:config}e. We observe that the profiles in the DTV reconstruction follow the truth profiles better than do those in the ITV reconstruction. Moreover, the DTV profiles from data over $\alpha_{\rm min}\ge 30^\circ$ coincide virtually completely with their corresponding truth profiles of the blurred-bar phantom.}

\begin{figure}[t]
\centering
\includegraphics[angle=0,trim=0 0 0 0, clip,origin=c,width=1.0\textwidth]{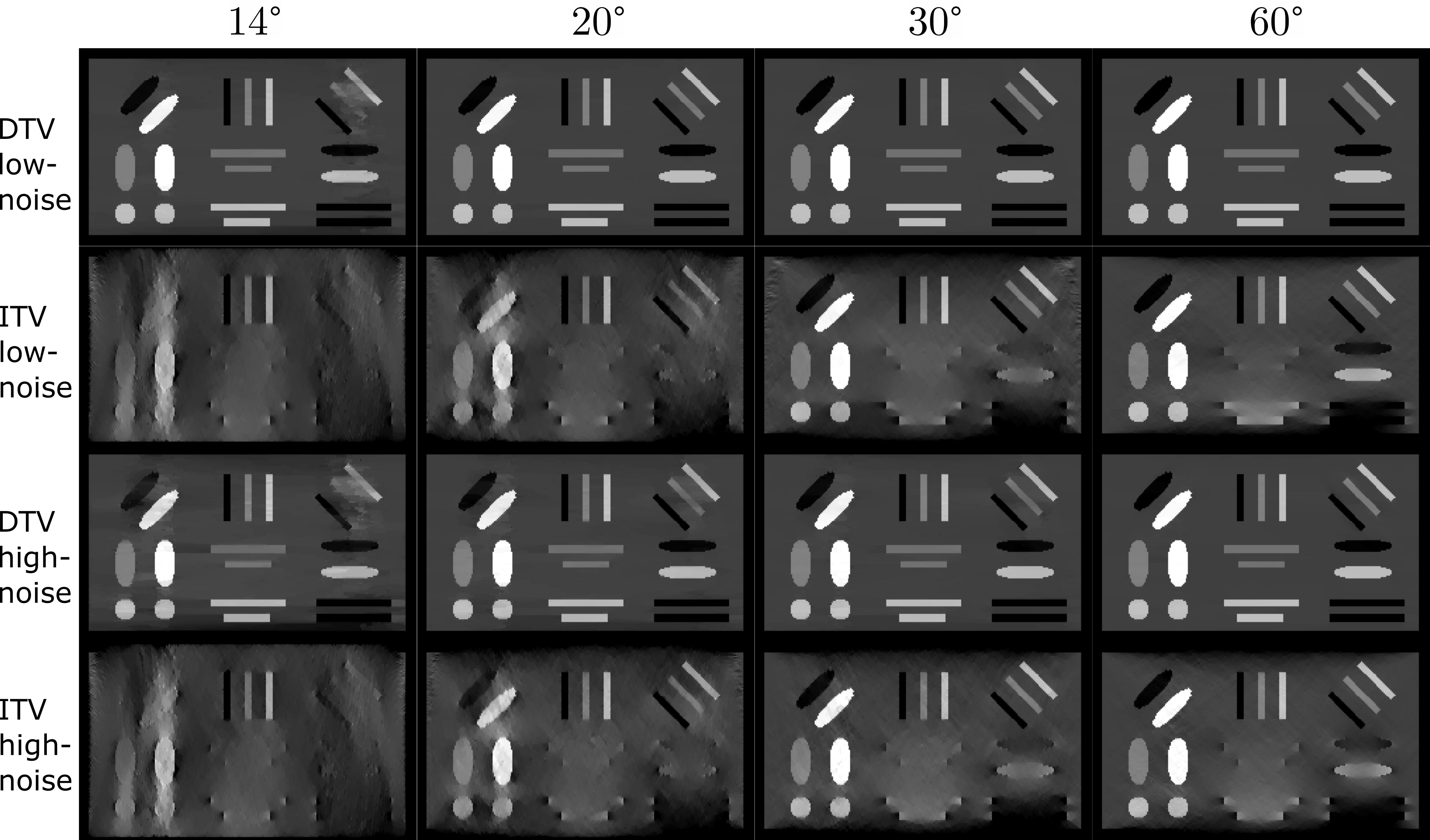}
\caption{{Images of the bar phantom reconstructed by use of the DTV (rows 1 $\&$ 3) and ITV  (rows 2 $\&$ 4) algorithms from  data of low- (rows 1 $\&$ 2) and high- (rows 3 $\&$ 4) noise levels over angular ranges of 14$^\circ$ (column 1), 20$^\circ$ (column 2), 30$^\circ$ (column 3), and 60$^\circ$ (column 4). Display window: [0.1, 0.5] cm$^{-1}$.}}
\label{fig:bar-angular-noisy}
\end{figure}

\subsection{Reconstruction of the bar phantoms from noisy limited-angular-range data } \label{sec:bar_noise}
We also perform a preliminary investigation of image reconstruction of the bar and blurred-bar phantoms from noisy, limited-angular-range data. Using noiseless data as the mean, which are obtained over one of the ten angular ranges described in Sec. \ref{sec:bar_angular_range} from one of the bar and blurred-bar phantoms, we generate noisy-data sets of multiple Poisson-noise levels and then perform reconstructions from the noisy-data sets. While we have performed studies for multiple data-noise levels for each of the ten angular ranges described above, we show reconstructions from data sets of low- and high-noise levels with SNRs of $\sim\!10^4$ and $\sim\!3.2\times10^3$, resulting in $\sim\!10^8$ and $\sim\!10^7$ emitting photons for each ray, only for four limited-angular ranges $\alpha = 14^\circ$, $20^\circ$, $30^\circ$, and $60^\circ$, respectively. Results for other angular ranges and noise levels are not included because observations similar to those presented below can be made based upon the results. 

\begin{figure}[t]
\centering
\includegraphics[angle=0,trim=0 0 0 0, clip,origin=c,width=1.0\textwidth]{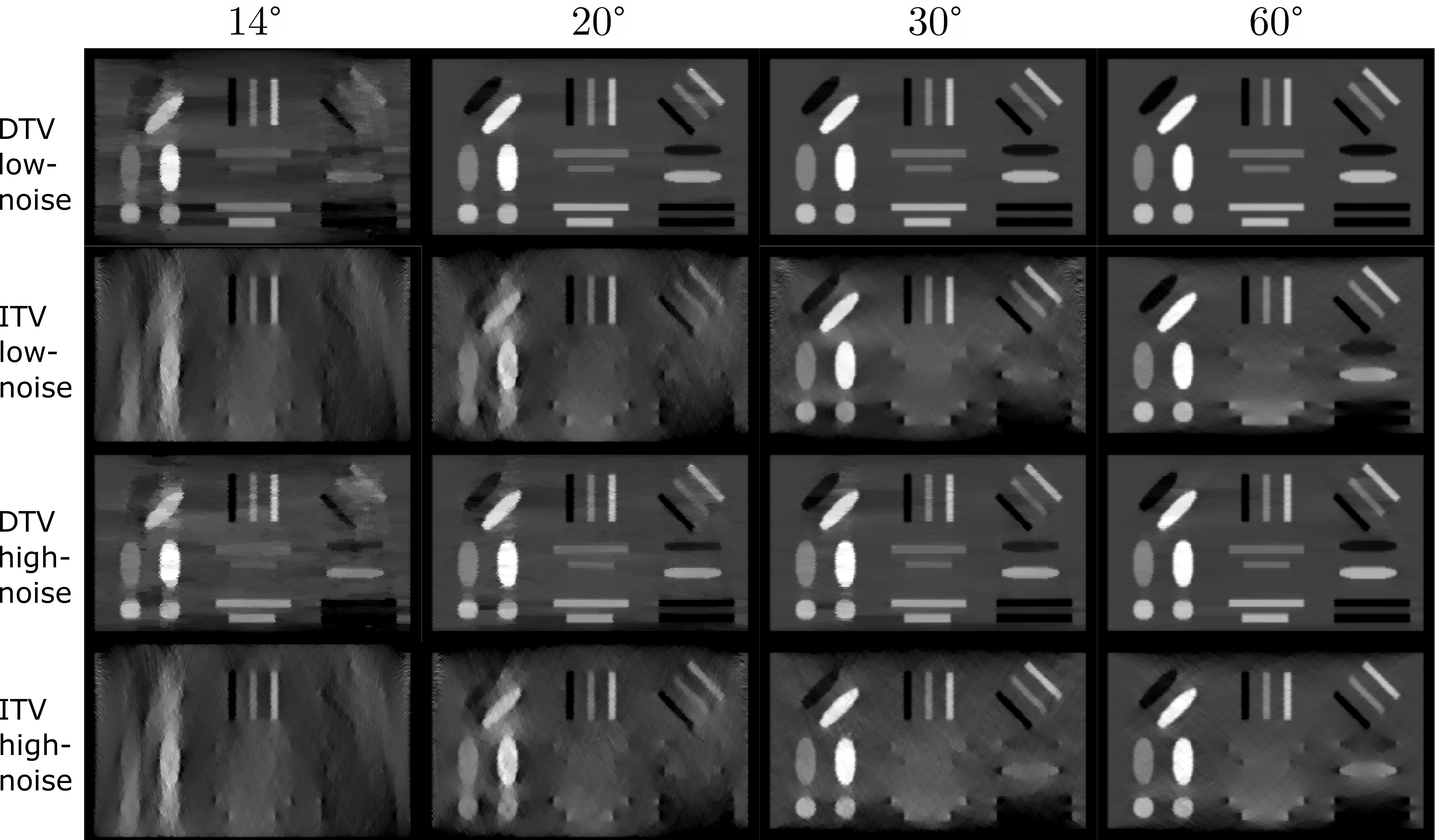}
\caption{{Images of the blurred-bar phantom reconstructed by use of the DTV (rows 1 $\&$ 3) and ITV  (rows 2 $\&$ 4) algorithms from data of low- (rows 1 $\&$ 2) and high- (rows 3 $\&$ 4) noise levels over angular ranges of 14$^\circ$ (column 1), 20$^\circ$ (column 2), 30$^\circ$ (column 3), and 60$^\circ$ (column 4). Display window: [0.1, 0.5] cm$^{-1}$.}}
\label{fig:blurred-bar-angular-noisy}
\end{figure}

Displaying images reconstructed for the bar phantom in Fig. \ref{fig:bar-angular-noisy}, we observe that DTV reconstructions in general show reduced artifacts comparing to the respective ITV reconstructions. The reconstruction quality diminishes as the angular range decreases from 60$^\circ$ to 14$^\circ$, and the lower the level of data noise the less artifacts are observed in the reconstructions.  Again,  reconstruction of the DTV algorithm is more robust than that of the ITV algorithm as the former contains visually less severe artifacts than does the latter for a given angular range. We have also computed nRMSEs of these reconstructions, which are not shown here, because observations can be obtained similar to those made for the nRMSE results above. 
In Fig. \ref{fig:blurred-bar-angular-noisy}, we display images of the blurred-bar phantom reconstructed from data of low- and high-noise levels for $\alpha=14^\circ$, $20^\circ$, $30^\circ$,  and $60^\circ$, respectively, by using the DTV and ITV algorithms.  Observations  can be made for the blurred-bar phantom similar to those for the study results of the bar phantom.

\section{Discussions}\label{sec:Discussion}




In this work, we investigate optimization-based image reconstruction from data over limited-angular ranges that are significantly smaller than the short-scan-angular range in CT.  Following the formulation of the reconstruction problem as a convex optimization program, we develop the DTV algorithm to solve the optimization program. Because it is the computer implementation of the DTV algorithm that is used in quantitative studies, we carry out a numerical study to verify the implementation correctness of the DTV algorithm by showing that it can solve numerically accurately the optimization program. Furthermore, under sufficient, consistent data conditions, the DTV algorithm can yield numerically accurate image reconstruction, i.e., inverting the DXT-data model.

It is of interest in obtaining an empirical condition on the minimal-angular range sufficient for yielding numerically accurate reconstructions for a class of objects. Clearly, such a minimal-angular range depends upon a number of factors, including object structure and the algorithm itself. Using the DTV algorithm, we perform empirical studies on image reconstructions from data generated over a number of limited-angular ranges for numerous phantoms  of practical relevance. Metrics are used for measuring quantitative accuracy of a reconstruction and its visual correlation to the truth image.  The quantitative study results suggest that for the study conditions considered,  the DTV algorithm may yield accurate reconstructions of the breast and bar phantoms from noiseless data generated over minimal-angular ranges  of $14^\circ\!\sim\! 30^\circ$, depending upon the complexity of the object structure, and that the minimal-angular ranges are considerably smaller than those obtained empirically with the ITV algorithm and other existing algorithms. 


Similar to any other algorithm, parameters are involved in the DTV reconstruction, such as DTV constraint parameters $t_x$ and $t_y$. In our simulation studies involving both noiseless and noisy data generated from a numerical phantom, we compute the DTVs of the truth image, i.e., the numerical phantom, and use them as $t_x$ and $t_y$. However, different selections of $t_x$ and $t_y$ can impact DTV reconstruction, as illustrated in \ref{sec:para_selection}. In a study in which knowledge of the truth image is absent, parameters $t_x$ and $t_y$ need to be determined empirically by, e.g., basing upon a quality or utility metric designed specifically for the study. One could perform multiple reconstructions with multiple sets of parameter values of $t_x$ and $t_y$, compute the metric value for each reconstruction obtained with a set of the parameter values, and choose the set that yields the empirically highest metric value. This is the approach that we have taken to determine parameters when an algorithm is applied in a real-data study in which the truth image is unknown  \citep{bian2014investigation,han2015algorithm,zhang2016investigation,xia2016optimization}. 

We have focused on image reconstruction for different limited-angular ranges with an identical angular interval between views. For a given limited-angular range, while it is expected that a reduced angular interval, i.e., an increased number of views, would improve reconstruction accuracy, it remains largely to be investigated as to the extent of such an improvement as a function of the angular interval and the associated angular range. The study may bear implication for optimally distributing scanning dose and time for a given imaging task and its workflow constraints.

{  The results of our numerical study suggest that the DTV algorithm may yield more accurate reconstruction than the ITV algorithm from limited-angular-range data. This is understandable as the two separate DTV constraints are likely to specify a feasible solution set tighter than that specified by a single ITV constraint. For the configuration described in Fig. \ref{fig:config}a, limited-angular-range artifacts appear as directional streaks largely along the $x$-axis in the FBP or ITV reconstruction \citep{quinto2017artifacts}.
In the DTV reconstruction, the individual constraints applied separately to image DTVs along $x$- and $y$-axes may allow for reconstruction of boundaries along the $y$-axis efficiently and subsequently help reconstruct the boundaries along $x$-axis. Conversely, the ITV constraint blends, and thus may destroy, the DTV-directional information, resulting in an ITV reconstruction with possibly more artifacts than the DTV algorithm when applied to limited-angular-range data.}

The optimization program in Eq. \eqref{eq:opt} includes data-$\ell_2$ norm and DTVs. It would be interesting to investigate additional designs of optimization programs and their associated algorithms for potentially further lowering the minimal-angular ranges obtained with the optimization program in Eq. \eqref{eq:opt} and its DTV algorithm. We are currently investigating to replace data-$\ell_2$ norm with data terms of different forms, including data-KL divergence and data-$\ell_1$ norm. As long as these new optimization programs remain convex, they can be solved accurately by use of the general PD algorithm. One may subsequently seek to derive and solve the proximal mappings corresponding to the optimization programs,  thus obtaining instances of the general PD algorithm for solving the optimization programs. The work focuses on reporting 2D-image reconstruction from limited-angular-range data. However, it can readily be extended to 3D-image reconstruction from data collected over limited-angular ranges. The ITV algorithm has been shown \citep{Sidky:08} to reduce the cone-beam artifacts observed in image reconstruction with analytic algorithms from circular cone-beam data. As the DTV algorithm is demonstrated in the work to be more effective than the ITV algorithm in accurate image reconstruction from limited-angular-range data, it would be interesting to investigate if the DTV algorithm  is more effective than the ITV algorithm in minimizing cone-beam artifacts in image reconstruction from circular cone-beam data collected especially over limited angular ranges. We are pursuing the extension and will be reporting the results in the near future.

\section{Conclusions}\label{sec:Conclusions}
In the work, we develop the DTV algorithm for image reconstruction from data collected over a limited-angular range that is substantially smaller than $\pi$ plus fan angle in a short-scan CT. The algorithm and its implementation achieve image reconstruction through solving an optimization program that includes DTV constraints. Using the DTV algorithm, we investigate image reconstructions from data generated over a number of limited-angular ranges of interest for breast and bar phantoms of potential relevance to clinical and industrial applications. Based upon the study results, we obtain empirically minimal-angular ranges sufficient for numerically accurately reconstructing images for scanning conditions and phantom anatomies considered. 

\section{Acknowledgment}
This work was supported in part by NIH R01 Grant Nos. EB026282, EB023968, and Grayson-Jockey Club Research. The computation of the work was performed in part on the computer cluster funded by NIH S10-OD025081, S10-RR021039,  and P30-CA14599 awards. The contents of this paper are solely the responsibility of the authors and do not necessarily represent the official views of NIH.

\newpage

\appendix
\addcontentsline{toc}{section}{Appendices}

\section{DTV algorithm}\label{sec:cpd}

\subsection{Derivation of the DTV algorithm}
It is well-known that the general PD algorithm \citep{rockafellar2015convex,c-p:2011} can solve the convex optimization program in Eq. \eqref{eq:opt}. However, the use of the general PD algorithm necessarily requires the computation of a proximal mapping in the algorithm; and direct numerical computation of the proximal mapping may result in diminished numerical accuracy and computational efficiency of the algorithm. For the optimization program given in Eq. \eqref{eq:opt}, we show below that a solution to its proximal mapping can indeed be derived. With that, we in essence obtain an instance of the general PD algorithm, which is referred to as the {DTV} algorithm. 

The derivation of the solution to the proximal mapping starts with the reformulation of the optimization program in Eq. \eqref{eq:opt} as
\begin{eqnarray}\label{eq:opt-1}
\mathbf{f}^{\star} & = & \underset{\mathbf{f}}{\mathsf{argmin}} \{ \frac{1}{2} \parallel \mathcal{H}\mathbf{f} - \mathbf{g}^{[\mathcal{M}]} \parallel^2_2 
		      + \delta_{{\rm Diamond}(\nu_1 t_x)}(\nu_1 |\mathcal{D}_x \mathbf{f}|_1)\nonumber \\
		&      + &  \delta_{{\rm Diamond}(\nu_2 t_y)}(\nu_2 |\mathcal{D}_y \mathbf{f}|_1) + \delta_P (\mu \mathbf{f}) \},
\end{eqnarray}
where indicator functions $\delta_{{\rm Diamond}(\beta)}(\mathbf{x})$ and $\delta_P (\mathbf{x})$ are defined as:
\begin{equation}
\delta_{{\rm Diamond}(\beta)}(\mathbf{x})= 
\begin{cases}
    0,&  ||\mathbf{x}||_1 \le \beta \\
    \infty,              & ||\mathbf{x}||_1 > \beta
\end{cases}, \quad\quad
\delta_P (\mathbf{x})= 
\begin{cases}
    0,&  \mathbf{x} \ge \mathbf{0}\\
    \infty,              & \text{Otherwise}
\end{cases}.
\end{equation}

We  consider a pair of primal and dual optimization problems, which can be solved by use of the general PD algorithm \citep{rockafellar2015convex,c-p:2011}: 
\begin{eqnarray}
\mathbf{x^{\star}} & = & \underset{\mathbf{x}}{\mathrm{argmin}}\left\{ F(\mathcal{K}\mathbf{x})+G(\mathbf{x})\right\} ,\label{eq:primal-prob}\\
\mathbf{y^{\star}} &=& \underset{\mathbf{y}}{\mathrm{argmax}}\left\{ -F^*(\mathbf{y})-G^*(-\mathcal{K}^{\top}\mathbf{y})\right\} ,\label{eq:dual-prob}
\end{eqnarray}
where $F$ and $G$ denote two convex functions, along with their respective convex conjugate functions $F^\ast$ and $G^\ast$, and $\mathcal{K}$ denotes a linear transform.  

We now design 
\begin{eqnarray}
\mathbf{x} &=&  \mathbf{f}, \quad \mathbf{r} = \mathcal{H}\mathbf{f}, \quad \mathbf{v} =  \nu_1 \mathcal{D}_x\mathbf{f}, \nonumber\\
\quad \mathbf{z} &=&  \nu_2 \mathcal{D}_y\mathbf{f}, \quad \mathbf{s} =  \mu \mathbf{f}, \quad
\mathcal{K}=  \left(\begin{array}{c}
\mathcal{H}\\
\nu_1\mathcal{D}_x\\
\nu_2\mathcal{D}_y\\
\mu \mathcal{I}
\end{array}\right),
\label{eq:assignment-variable}
\end{eqnarray}
and
\begin{eqnarray}
F(\mathcal{K} \mathbf{x}) &=& F(\mathbf{r},\mathbf{v},\mathbf{z},\mathbf{s}) = F_{1}(\mathbf{r})+F_{2}(\mathbf{v})+F_{3}(\mathbf{z})+F_{4}(\mathbf{s}), \label{eq:assignment-F} \\
F_{1}(\mathbf{r}) & = & \frac{1}{2} \parallel \mathbf{r} - \mathbf{g}^{[\mathcal{M}]} \parallel ^2_2 \label{eq:assignment-f1} \\
F_{2}(\mathbf{v}) & = & \delta_{{\rm Diamond}(\nu_1 t_x)}(|\mathbf{v}|)  \label{eq:assignment-f2} \\
F_{3}(\mathbf{z}) & = & \delta_{{\rm Diamond}(\nu_2 t_y)}(|\mathbf{z}|) \label{eq:assignment-f3} \\
F_{4}(\mathbf{s}) & = & \delta_{P}(\mathbf{s}) \label{eq:assignment-f4} \\
G(\mathbf{x}) & = & 0.\label{eq:assignment-G}
\end{eqnarray}
Considering Eqs. (\ref{eq:assignment-F})-(\ref{eq:assignment-G}), we can 
rewrite Eq. \eqref{eq:primal-prob} as a primal minimization:
\begin{eqnarray}\label{eq:primal-new}
\mathbf{f}^{\star}
 & = & \underset{\mathbf{f}}{\mathrm{argmin}}\{ \frac{1}{2} \parallel \mathbf{r} - \mathbf{g}^{[\mathcal{M}]} \parallel _2^2
                                                   + \delta_{{\rm Diamond}(\nu_1 t_x)}(| \mathbf{v}|)\nonumber \\
                                                &    + & \delta_{{\rm Diamond}(\nu_2 t_y)}(|\mathbf{z}|) +\delta_{P}(\mathbf{s})\}.
\end{eqnarray}

Conversely, the convex conjugate functions $F^\ast$ and $G^\ast$ can be obtained as
\begin{align}
F^{*}(\mathbf{w},\mathbf{p},\mathbf{q},\mathbf{t})
=&  F_{1}^{*}(\mathbf{w})+F_{2}^{*}(\mathbf{p})+F_{3}^{*}(\mathbf{q})+F_ {4}^{*}(\mathbf{t}) 
\nonumber \\
G^{*}(\mathbf{y}) 
  =& \delta_{0}(\mathbf{y}),\label{eq:conjugate-G} 
\end{align}
where
\begin{eqnarray}
F_{1}^{*}(\mathbf{w}) 
 & = & \frac{1}{2} \parallel \mathbf{w} \parallel ^2 + \mathbf{w}^{\top}\mathbf{g}^{[\mathcal{M}]}, \label{eq:conjugate-f1} \\
F_{2}^{*}(\mathbf{p})  & = & \nu_1 t_x ||(|\mathbf{p}|)||_{\infty}, \label{eq:conjugate-f2} \\
F_{3}^{*}(\mathbf{q})  & = &  \nu_2 t_y ||(|\mathbf{q}|)||_{\infty},\label{eq:conjugate-f3}\\
F_{4}^{*}(\mathbf{t})  & = &  \delta_{P}(-\mathbf{t}),\label{eq:conjugate-f4}
\end{eqnarray}
$ ||\cdot||_{\infty}$ denotes the largest entry of the vector, and function $\delta_{0}(\mathbf{y})$ is defined as
\begin{equation}
\delta_0 (\mathbf{y})= 
\begin{cases}
    0,&  \mathbf{y} = \mathbf{0}\\
    \infty,              & \text{Otherwise}
\end{cases}. \nonumber
\end{equation} Substituting Eqs. (\ref{eq:conjugate-G})-(\ref{eq:conjugate-f4}) into Eq. (\ref{eq:dual-prob})
and noticing $\mathcal{K}^\top=(\mathcal{H}^{\top},\nu_1\mathcal{D}_x^{\top}, \nu_2\mathcal{D}_y^{\top}, \mu \mathcal{I})$, we obtain the dual maximization problem as
\begin{eqnarray}
\hspace{-.cm} (\mathbf{w}^{\star},\mathbf{p}^{\star},\mathbf{q}^{\star},\mathbf{t}^{\star}) & = &
\underset{\mathbf{w},\mathbf{p},\mathbf{q},\mathbf{t}}{\rm argmax}\{
 - \frac{1}{2} \parallel \mathbf{w} \parallel ^2 - \mathbf{w}^{\top}\mathbf{g}^{[\mathcal{M}]}  - \nu_1 t_x ||(|\mathbf{p}|)||_{\infty}  -\nu_2 t_y ||(|\mathbf{q}|)||_{\infty}\nonumber \\
 & - &  
 \delta_P(-\mathbf{t})- \delta_0 (-\mathcal{H}^{\top}\mathbf{w} - \nu_1\mathcal{D}_x^{\top} \mathbf{p} - \nu_2\mathcal{D}_y^{\top} \mathbf{q} - \mu \mathbf{t}) \}.\label{eq:pdgap}
\end{eqnarray}

%

When the general PD algorithm \citep{rockafellar2015convex,c-p:2011} is used to solve mathematically exactly the pair of primal and dual optimization problems in Eqs. \eqref{eq:primal-new} and \eqref{eq:pdgap}, it also automatically solves our optimization program in Eq. \eqref{eq:opt} as it is identical to the primal optimization in Eq. \eqref{eq:primal-new}. The proximal mapping is used to generate a descent direction for solving the PD problem. We take functions $F_{1}^{*}(\mathbf{w})$ and $G(\mathbf{x})$ as an example, and their proximal mapping ${\rm prox}_{\sigma}[F_{1}^{*}](\mathbf{w})$ and ${\rm prox}_{\tau}[G](\mathbf{x})$ are defined as:
\begin{equation}\label{eq:proxF1}
{\rm prox}_{\sigma}[F_{1}^{*}](\mathbf{w})  =  \underset{\mathbf{w}^{\prime}}{\mathsf{argmin}} \{ F_{1}^*(\mathbf{w}^\prime) + \frac{||\mathbf{w}-\mathbf{w}^\prime||^2_2}{2\sigma} \},
\end{equation}
\begin{equation}\label{eq:G}
{\rm prox}_{\tau}[G](\mathbf{x})  =  \underset{\mathbf{x}^{\prime}}{\mathsf{argmin}} \{ G(\mathbf{x}^\prime) + \frac{||\mathbf{x}-\mathbf{x}^\prime||^2_2}{2\tau} \}.
\end{equation}
As shown below, we derive analytical solutions to the proximal mappings and thus obtain the DTV algorithm as an instance of the general PD algorithm specific to  optimization program in Eq. \eqref{eq:opt}.

Using Eqs. \eqref{eq:proxF1}-\eqref{eq:G}, we can readily obtain the analytical solutions of proximal mappings ${\rm prox}_\sigma [F^*]$ and ${\rm prox}_\sigma [G]$ as 
\begin{eqnarray}
\hspace{-1.5cm}
{\rm prox}_{\sigma}[F^{*}](\mathbf{w},\mathbf{p},\mathbf{q},\mathbf{t}) & = & {\rm prox}_{\sigma}[F_{1}^{*}](\mathbf{w}) + {\rm prox}_{\sigma}[F_{2}^{*}](\mathbf{p}) \nonumber\\
&+& {\rm prox}_{\sigma}[F_{3}^{*}](\mathbf{q}) + {\rm prox}_{\sigma}[F_{4}^{*}](\mathbf{t})\nonumber\\
\hspace{-1.5cm}{\rm prox}_{\tau}[G](\mathbf{x}) & = & \mathbf{x},\label{eq:proximalG}
\label{eq:proximal-F-G}
\end{eqnarray}
where 
\begin{eqnarray}
{\rm prox}_{\sigma}[F_{1}^{*}](\mathbf{w}) & = & \frac{\mathbf{w} - \sigma \mathbf{g}^{[\mathcal{M}]}}{1+\sigma}
\label{eq:proximal-f1}\\
{\rm prox}_{\sigma}[F_{2}^{*}](\mathbf{p})  & = & \mathbf{p} - \sigma \mathbf{p} \ell_1 {\rm ball}_{\nu_1 t_x} (|\mathbf{p}|/\sigma)/|\mathbf{p}|
\label{eq:proximal-f2}\\
{\rm prox}_{\sigma}[F_{3}^{*}](\mathbf{q}) & = & \mathbf{q} - \sigma \mathbf{q} \ell_1 {\rm ball}_{\nu_2 t_y} (|\mathbf{q}|/\sigma)/|\mathbf{q}|
\label{eq:proximal-f3}\\
{\rm prox}_{\sigma}[F_{4}^{*}](\mathbf{t})  & =  & {\rm neg}(\mathbf{t}),\label{eq:proximalF4}
\label{eq:proximal-f4}
\end{eqnarray}
which are obtained with Eqs. \eqref{eq:conjugate-f1}-\eqref{eq:conjugate-f4}. Plugging the analytical results  of Eqs. \eqref{eq:proximalG}-\eqref{eq:proximalF4} into the pseudo-code of the general PD algorithm \citep{c-p:2011,sidky2012convex}, we thus obtain the pseudo-code in Sec. \ref{sec:DTV} for the DTV algorithm. 


\subsection{Convergence conditions on the DTV algorithm}\label{sec:convergence}

We first devise four convergence conditions for both consistent and inconsistent data   for the DTV algorithm as \citep{sidky2012convex,chen2021non}
\begin{equation}\label{eq:convergence_1}
d\widetilde{D}^{(n)}_\mathbf{g}\rightarrow 0,  \,\,\,\,\,\,  \widetilde{D}^{(n)}_{\rm TV_{\it x}}\rightarrow 0,  \,\,\,\,\,\,  \widetilde{D}^{(n)}_{\rm TV_{\it y}}\rightarrow 0,
\,\,\,\,\,\, d\widetilde{D}^{(n)}_\mathbf{f}\rightarrow 0,
\end{equation}
as $n\rightarrow \infty$, in which the dimensionless metrics are defined as:
\begin{equation}\label{eq:convergence-formula-1}
\begin{split}
d\widetilde{D}^{(n)}_\mathbf{g} &= |\sqrt{D_{\mathbf{g}}(\mathbf{f}^{(n)})}
 - \sqrt{D_{\mathbf{g}}(\mathbf{f}^{(n-1)})}|/||\mathbf{g}^{[\mathcal{M}]}||_2 \\
 \widetilde{D}^{(n)}_{\rm TV_{\it x}} &= | (||(|\mathcal{D}_x \mathbf{f}^{(n)}|)||_1 - t_x)|/t_x \\
 \widetilde{D}^{(n)}_{\rm TV_{\it y}} &= | (||(|\mathcal{D}_y \mathbf{f}^{(n)}|)||_1 - t_y)|/t_y \\
 d\widetilde{D}^{(n)}_\mathbf{f} &= ||\mathbf{f}^{(n)}-\mathbf{f}^{(n-1)}||_2/||\mathbf{f}^{(n-1)}||_2. \\
\end{split}
\end{equation}

Furthermore, we design three additional convergence conditions for both consistent and inconsistent data as \citep{sidky2012convex,chen2021non}
\begin{equation}\label{eq:convergence_2}
\widetilde{\rm cPD}^{(n)} \rightarrow 0,  \quad  \widetilde{\rm T}^{(n)} \rightarrow 0,  \quad \widetilde{\rm S}^{(n)} \rightarrow 0, 
\end{equation}
as $n\rightarrow \infty$, in which the dimensionless metrics are give by
\begin{equation}\label{eq:convergence_2a}
\widetilde{\rm cPD}^{(n)} =\frac{{\rm cPD}^{(n)}}{{\rm cPD}^{(1)}},  \,\,\,\,\,\,  \widetilde{\rm T}^{(n)} =\frac{{\rm T}^{(n)}}{{\rm T}^{(1)}},  \,\,\,\,\,\, \widetilde{\rm S}^{(n)} =\frac{{\rm S}^{(n)}}{{\rm S}^{(1)}},  
\end{equation}
with conditional primal-dual (cPD) gap cPD$^{(n)}$ \citep{xia2016optimization}, transversality T$^{(n)}$ \citep{hiriart2013convex}, and dual gap S$^{(n)}$ \citep{goldstein2013adaptive} given by
\begin{equation}\label{eq:convergence-formula-2}
\begin{split}
{\rm cPD}^{(n)} &= \frac{1}{2}||\mathcal{H}\mathbf{f}^{(n)}-\mathbf{g}^{[\mathcal{M}]}||_2^2 + \frac{1}{2} \parallel \mathbf{w}^{(n)} \parallel_2 ^2 + \mathbf{w}^{(n)\top}\mathbf{g}^{[\mathcal{M}]}  \\
&+ \nu_1 t_x ||(|\mathbf{p^{(n)}}|)||_{\infty}  + \nu_2 t_y ||(|\mathbf{q^{(n)}}|)||_{\infty} \\
{\rm T}^{(n)} &= ||\mathcal{H}^{\top}\mathbf{w}^{(n)}+\nu_1\mathcal{D}_x\mathbf{p}^{(n)}+\nu_2\mathcal{D}_y\mathbf{q}^{(n)} + \mu \mathbf{t}^{(n)}||_2 \\
{\rm S}^{(n)} &= ||\frac{1}{\sigma}\begin{pmatrix} 
\mathbf{w}^{(n)}-\mathbf{w}^{(n-1)}\\
\mathbf{p}^{(n)}-\mathbf{p}^{(n-1)}\\
\mathbf{q}^{(n)}-\mathbf{q}^{(n-1)}\\
\mathbf{t}^{(n)}-\mathbf{t}^{(n-1)}
\end{pmatrix}
-\begin{pmatrix} 
\mathcal{H}\\
\nu_1 \mathcal{D}_x\\
\nu_2 \mathcal{D}_y\\
\mu \mathcal{I}
\end{pmatrix}(\mathbf{f}^{(n)}-\mathbf{f}^{(n-1)})
||_2.\\
\end{split}
\end{equation}
In the numerical studies in the work, reconstructions are obtained when convergence conditions in Eqs. \eqref{eq:convergence_1} and \eqref{eq:convergence_2} are achieved numerically, as the example in Sec. \ref{sec:Verify} below demonstrates.

Clearly, data divergence ${D_{\mathbf{g}}(\mathbf{f})}$ itself can also be used to form a necessary convergence condition as  
\begin{equation}\label{eq:ddist}
\widetilde{D}^{(n)}_\mathbf{g} \rightarrow c\,\,\,\,\,\,\, {\rm with}\,\,\,\,\,\,\, 
\widetilde{D}^{(n)}_\mathbf{g} = |\sqrt{D_{\mathbf{g}}(\mathbf{f}^{(n)})}|/||\mathbf{g}^{[\mathcal{M}]}||_2,
\end{equation}
as  $n\rightarrow \infty$,
where $c$ is a non-negative constant satisfying $c=0$ for consistent data and $c >0$ for inconsistent data such as real data.

Convergence conditions are a part of the DTV algorithm, and each of the multiple convergence conditions discussed above
characterizes an aspect of the convergence property of the algorithm. The metrics used for devising the convergence conditions are normalized so that the conditions are dimensionless and fall into roughly comparable numerical ranges.

\section{Numerical verification of the DTV algorithm}\label{sec:Verify}

Without loss of generality, we present here a verification study in which noiseless data are acquired from the breast phantom over a full-angular range of $2\pi$ with an angular interval of $1^\circ$ by use of the scanning configuration in Fig. \ref{fig:config}a. A leading reason to consider a full-angular range of $2\pi$ is to, in addition to show convergence curves of Eqs. \eqref{eq:convergence_1} and \eqref{eq:convergence_2}, demonstrate that the DTV algorithm can also, under sufficient, consistent data condition, numerically accurately recover $\mathbf{f}^{[\rm truth]}$ by inverting  the DXT-data model in Eq. \eqref{eq:DXT}.

\begin{figure}
\centering
\includegraphics[angle=0,trim=0 0 0 0, clip,origin=c,width=1.\textwidth]{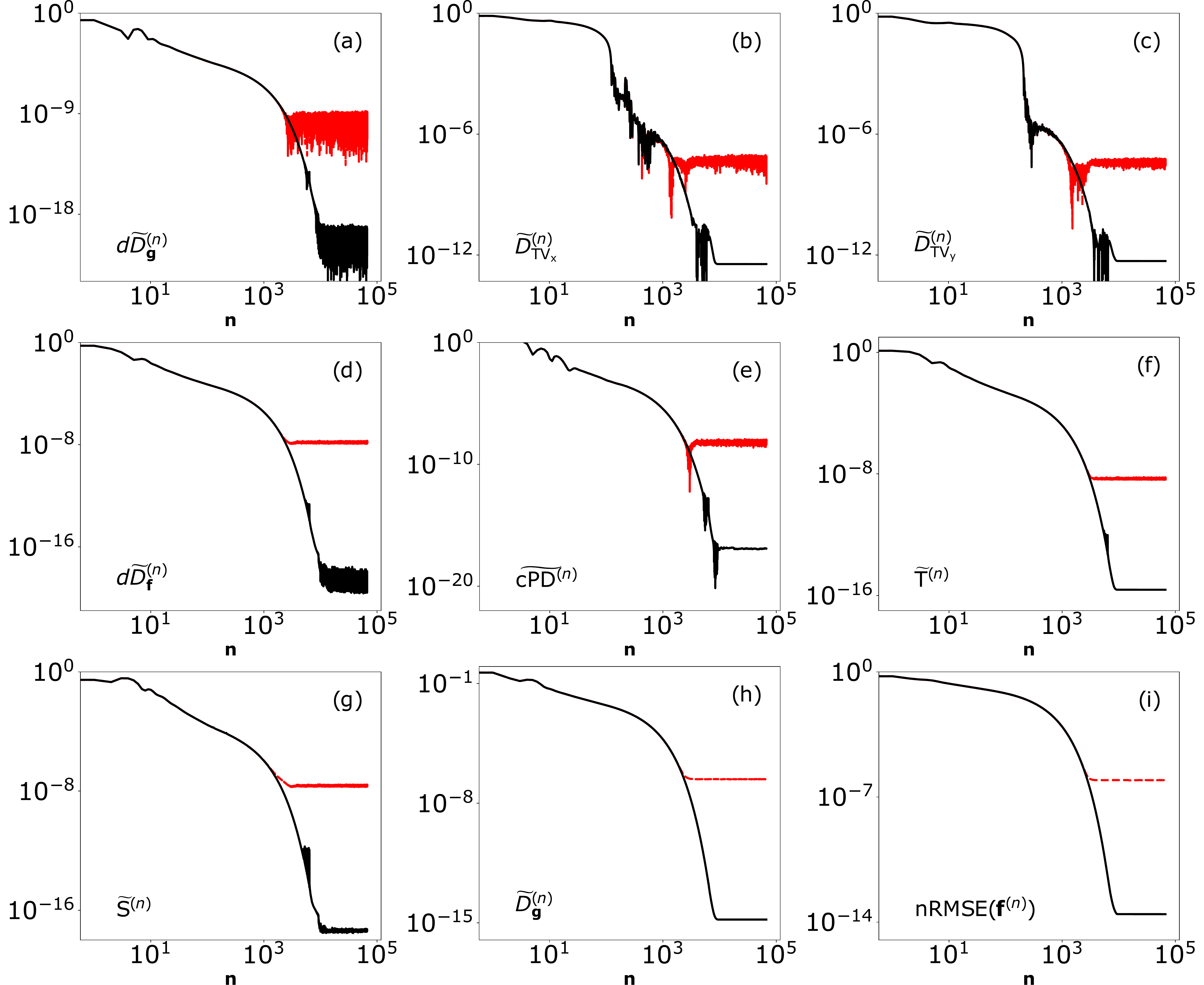}
\caption{{Convergence metrics (a) $d\widetilde{D}^{(n)}_\mathbf{g}$, (b) $\widetilde{D}^{(n)}_{\rm TV_{\it x}}$, (c) $\widetilde{D}^{(n)}_{\rm TV_{\it y}}$, (d) $d\widetilde{D}^{(n)}_\mathbf{f}$, (e) $\widetilde{\rm cPD}^{(n)}$, (f) $\widetilde{\rm T}^{(n)}$, (g) $\widetilde{\rm S}^{(n)}$, (h) $\widetilde{D}^{(n)}_\mathbf{g}$, and (i) ${\rm nRMSE}(\mathbf{f}^{(n)})$ of the DTV algorithm, as functions of iteration $n$ obtained with single (dashed, red) and double (solid, black) floating-point computer precision.}}
\label{fig:verification-plots}
\end{figure}
We apply the DTV algorithm to reconstructing images from noiseless (i.e., consistent) full-angular-range data and then compute the convergence metrics in Eqs. \eqref{eq:convergence_1}, \eqref{eq:convergence_2}, and \eqref{eq:ddist}, thus obtaining a total of eight convergence curves shown in Figs. \ref{fig:verification-plots}a-\ref{fig:verification-plots}h in which ashed, red and solid, black curves denote results obtained with 4-byte (i.e., single) and 8-byte (i.e., double) floating-point precision. The convergence curves (dashed, red) obtained with the single floating-point precision in Fig. \ref{fig:verification-plots} decay until around 1000 iterations. In an attempt to demonstrate that their decays are limited largely by the computer precision, we conduct the same study by using double floating-point precision and obtained the convergence curves (solid, black) in Fig. \ref{fig:verification-plots}. It can be observed that the convergence curves obtained with double floating-point precision continue to decay beyond 1000 iterations until $\sim$10000 iterations. These curves are plotted in log-log scales for revealing unambiguously not only convergence metric values, but also equally importantly their decaying trends before approaching the computer precision.


\subsection{Metric for measuring numerical accuracy of reconstruction}

\begin{figure}
\centering
\includegraphics[angle=0,trim=0 0 0 0, clip,origin=c,width=1.\textwidth]{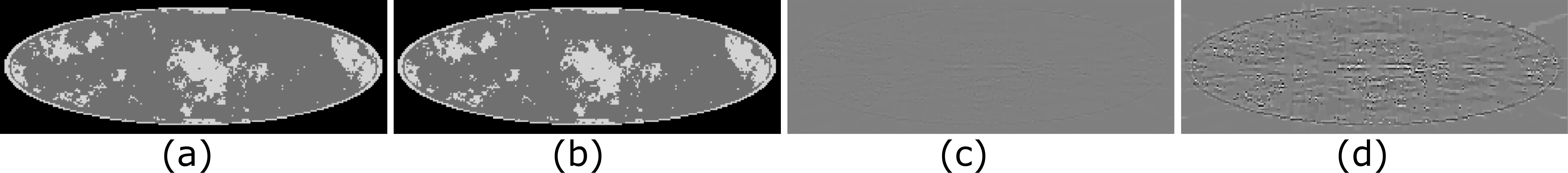}
\caption{{(a) Breast phantom, (b) reconstructed image, and (c) $\&$ (d) differences between truth and reconstructed images. Display window: [0.15, 0.25] cm$^{-1}$ for (a) and (b), [-$10^{-5}$, $10^{-5}$] cm$^{-1}$ for (c), and [-$10^{-6}$, $10^{-6}$] cm$^{-1}$ for (d).}}
\label{fig:breast-inverseCrime}
\end{figure}

While the results in Figs. \ref{fig:verification-plots}a-\ref{fig:verification-plots}h numerically verify the DTV algorithm and its computer implementation in terms of solving the optimization program in 
Eq. \eqref{eq:opt}, it remains to show if the DTV algorithm can, under sufficient, consistent data condition (i.e., $\mathbf{g}^{[\mathcal{M}]}=\mathcal{H}\mathbf{f}^{[\rm truth]}$, where $\mathbf{f}^{[\rm truth]}$ is the breast phantom) invert the DXT-data model, or equivalently, obtain $\mathbf{f}^{(n)} \rightarrow \mathbf{f}^{\rm [truth]}$ as $n \rightarrow \infty$. In an attempt to verify this, we use  the metric below to measure the inversion accuracy of the DXT-data model:
\begin{equation}\label{eq:imdist}
{\rm nRMSE}(\mathbf{f}^{(n)}) = ||\mathbf{f}^{(n)}-\mathbf{f}^{\rm [truth]}||_2/||\mathbf{f}^{\rm [truth]}||_2 \rightarrow 0,
\end{equation}
where ${\rm nRMSE}(\mathbf{f}^{(n)})$ denotes the normalized RMSE between the truth and reconstructed images, and then compute it with single and double floating-point precision, which are displayed as dashed, red and solid, black curves in Fig. \ref{fig:verification-plots}i. It can be observed that, as the iteration number $n$ increases, ${\rm nRMSE}(\mathbf{f})$ continues to decay until it achieves the computer precision. As displayed in Fig. \ref{fig:breast-inverseCrime}, the corresponding converged reconstruction is visually and numerically identical to the breast phantom, verifying that the DTV algorithm can, under sufficient consistent data condition, numerically accurately invert the DXT-data model.

\subsection{Metrics for measuring visualization accuracy of reconstruction}

We also use two additional metrics, i.e., the  the Pearson correlation coefficient (PCC) and normalized mutual information (nMI) \citep{pearson1895notes,Viergever:2003,Bian-PMB:2010},  to measure visualization correlation between reconstruction $\mathbf{f}^{(n)}$ and reference image $\mathbf{f}^{\rm [ref]}$. In the work, $\mathbf{f}^{\rm [ref]}=\mathbf{f}^{\rm [truth]}$. Metric PCC is given by
\begin{eqnarray}\label{eq:PCC}
\hspace{-1.5cm}
{\rm PCC}(\mathbf{f}^{(n)}) & = & \frac{\vert \rm Cov(\mathbf{f}^{(n)}, \mathbf{f}^{\rm [ref]})\vert}{\sigma(\mathbf{f}^{(n)}) \, \sigma(\mathbf{f}^{\rm [ref]})}
\end{eqnarray}
where ${\rm Cov}(\mathbf{f}^{(n)}, \mathbf{f}^{\rm [ref]})$ denotes the covariance between $\mathbf{f}^{(n)}$ and $\mathbf{f}^{\rm [ref]}$, i.e.,
\begin{eqnarray}\label{eq:cov-var}
 {\rm Cov}(\mathbf{f}^{(n)}, \mathbf{f}^{\rm [ref]}) 
&=& \frac{1}{N-1} \sum_{i=0}^{N-1} (f^{(n)}_i - \frac{1}{N}\sum_{i'=0}^{N-1}{f}^{(n)}_{i'})(f^{\rm [ref]}_i - \frac{1}{N}\sum_{i'=0}^{N-1}f^{\rm [ref]}_{i'}), \nonumber
\end{eqnarray}
and $\sigma^{2}(\mathbf{f}^{(n)})={\rm Cov}(\mathbf{f}^{(n)}, \mathbf{f}^{(n)})$ and $\sigma^{2}(\mathbf{f}^{\rm [ref]})={\rm Cov}(\mathbf{f}^{\rm [ref]}, \mathbf{f}^{\rm [ref]})$ indicate the variances of $\mathbf{f}^{(n)}$ and $\mathbf{f}^{\rm [ref]}$. Note that $0\le {\rm PCC}(\mathbf{f}^{(n)}) \le 1$. 
 
\noindent On the other hand, metric nMI is defined as
\begin{eqnarray}\label{eq:nMI}
\hspace{-1.5cm}{\rm nMI}(\mathbf{f}^{(n)}) & = & \frac{{\rm MI}(\mathbf{f}^{(n)})}{{\rm MI}(\mathbf{f}^{\rm [ref]})},
\end{eqnarray}
where  MI denotes mutual information between $\mathbf{f}^{(n)}$ and $\mathbf{f}^{\rm [ref]}$, given by
\begin{eqnarray}\label{eq:MI}
\hspace{-1.5cm}{\rm MI}(\mathbf{f}^{(n)}) & = & \sum_{i=0}^{N-1} \sum_{i^{\prime}=0}^{N-1}{\rm p}(f^{(n)}_i,f^{\rm [ref]}_{i^{\prime}}) \,{\rm log}\!
								  \left[ \frac{{\rm p}(f^{(n)}_i, \,f^{\rm [ref]}_{i^{\prime}})}{{\rm p}(f^{(n)}_i)\, {\rm p}(f^{\rm [ref]}_{i^{\prime}})} \right],\nonumber
\end{eqnarray}
${\rm p}(f^{(n)}_i)$ and ${\rm p}(f^{\rm [ref]}_{i})$ denote the ``marginal densities'' calculated from histograms of $\mathbf{f}^{(n)}$ and $\mathbf{f}^{\rm [ref]}$, and ${\rm p}(f^{(n)}_i, f^{\rm [ref]}_{i^{\prime}})$ depicts the ``joint density'' calculated from a 2D joint histogram of $\mathbf{f}^{(n)}$ and $\mathbf{f}^{\rm [ref]}$. 
Note that $0\le {\rm nMI}(\mathbf{f}^{(n)}) \le 1$.

Metrics PCC and nMI measure the degree of visual correlation between the reconstruction and reference images. In particular, the closer to 1  the PCC and nMI are, the more the reconstruction visually resembles the reference image, i.e., the less visual artifacts the reconstruction contains. 

{ 
\section{Impact of DTV-constraint parameters $t_x$ and $t_y$ on reconstruction}\label{sec:para_selection}

We discuss below how DTV-constraint parameters  $t_x$ and $t_y$ impact image reconstruction. Without loss of generality, we consider image reconstruction from consistent data, and let  $t_{x0}$  and $t_{y0}$ denote the DTVs of the truth image. In  Eq. \eqref{eq:opt}, data distance $D_{\mathbf{g}}(\mathbf{f})=0$ specifies a convex solution set, denoted as $\mathcal{B}$, which necessarily contains the truth image, whereas for a pair of selected  $t_x$ and $t_y$ in  Eq. \eqref{eq:opt}, DTV constraints $|| (|\mathcal{D}_x\mathbf{f}|) ||_1 \le t_x$ and $|| (|\mathcal{D}_y\mathbf{f}|) ||_1 \le t_y$ specify a convex solution set, denoted as $\mathcal{A}_{\rm DTV}$. We also use $\mathcal{A}_{\rm DTV}\cap\mathcal{B}$ to denote the intersection of sets $\mathcal{A}_{\rm DTV}$ and $\mathcal{B}$.

If one chooses $t_x<t_{x0}$ or $t_y<t_{y0}$, neither $\mathcal{A}_{\rm DTV}$ nor $\mathcal{A}_{\rm DTV}\cap\mathcal{B}$ contains the truth image. Therefore, images reconstructed by the DTV algorithm with  $t_x<t_{x0}$ or $t_y<t_{y0}$ are always different from the truth image. If one chooses $t_x\! \ge \! t_{x0}\,\&\,t_y\!>\!t_{y0}$ or $t_x\!>\!t_{x0}\,\&\,t_y\! \ge \! t_{y0}$, the truth image is interior to $\mathcal{A}_{\rm DTV}$ and also to $\mathcal{A}_{\rm DTV}\cap\mathcal{B}$. Therefore, $\mathcal{A}_{\rm ITV}\cap\mathcal{B}$ may contain more than one image, and DTV images obtained are likely to be different from the truth image. If one chooses $t_x=t_{x0}\,\&\,t_y=t_{y0}$, $\mathcal{A}_{\rm DTV}$ and thus $\mathcal{A}_{\rm DTV}\cap\mathcal{B}$ are the tightest sets containing the truth image; and if $\mathcal{A}_{\rm DTV}\cap\mathcal{B}$ contains only a single image, it is necessarily the truth image because we already know that the truth image is in $\mathcal{A}_{\rm DTV}\cap\mathcal{B}$.

\begin{figure}
\centering
\includegraphics[angle=0,trim=0 0 0 0, clip,origin=c,width=0.99\textwidth]{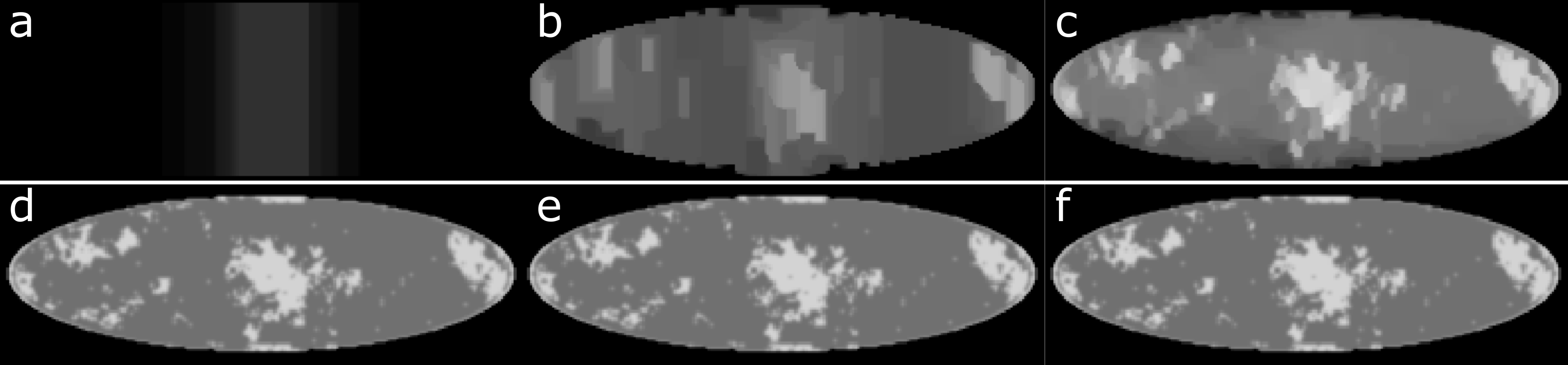}
\caption{{  Images of the blurred-breast phantom reconstructed from data over an angular range of $120^\circ$ by use of the DTV algorithm with $(t_x, t_y)=(0.5t_{x0}, 0.5t_{y0})$ (a), $(0.75t_{x0}, 0.75t_{y0})$ (b), $(0.85t_{x0}, 0.85t_{y0})$ (c), $(t_{x0}, t_{y0})$ (d),  $(1.25t_{x0}, 1.25t_{y0})$ (e), and $(1.5t_{x0}, 1.5t_{y0})$  (f), where $t_{x0}$ and $t_{y0}$ denote DTVs of the blurred-breast phantom. Display window: [0.15, 0.25] cm$^{-1}$.}}
\label{fig:blurred_breast_dtv_t}
\end{figure}

In an attempt to illustrate the impact of DTV constraint parameters on reconstruction, we conduct image reconstructions of the blurred-breast phantom from data collected over an angular range of $120^\circ$ by use of the DTV algorithm with multiple selections of  $t_x$ and $ t_y$, relative to truth values $t_{x0}$ and $t_{y0}$. 
In the top row of Fig. \ref{fig:blurred_breast_dtv_t}, we display images reconstructed with $(t_x, t_y)=(0.5t_{x0}, 0.5t_{y0})$, $(0.75t_{x0}, 0.75t_{y0})$, and $(0.85t_{x0}, 0.85t_{y0})$, respectively. It can be observed that these reconstructions clearly differ from the truth image, i.e., the blurred-breast phantom, because they are obtained with $t_x<t_{x0}$ or $t_y<t_{y0}$, as discussed above. On the other hand, 
the image in  Fig. \ref{fig:blurred_breast_dtv_t}d obtained with $(t_x, t_y)=(t_{x0}, t_{y0})$ is numerically identical to the truth image, i.e., the blurred-breast phantom, because $(t_x, t_y)=(t_{x0}, t_{y0})$ yields the tightest $\mathcal{A}_{\rm DTV}\cap\mathcal{B}$ that contains the truth image, as discussed above. Images shown in Figs. \ref{fig:blurred_breast_dtv_t}e and \ref{fig:blurred_breast_dtv_t}f are obtained with $(t_x, t_y)=(1.25t_{x0}, 1.25t_{y0})$ and $(1.5t_{x0}, 1.5t_{y0})$. While they appear visually similar to the truth image, they differ numerically from the truth image, and the extent of their difference depends upon the difference extent between $(t_x, t_y)$ used and $(t_{x0}, t_{y0})$ that determines the tightness of $\mathcal{A}_{\rm DTV}$.



}
\bibliography{ctrecon}
%

\end{document}